\documentclass[10pt]{iopart}

\usepackage{iopams}
\expandafter\let\csname equation*\endcsname\relax

\expandafter\let\csname endequation*\endcsname\relax

\usepackage{amsmath}

\usepackage{stmaryrd} 

\usepackage{MnSymbol} 
\usepackage{csquotes}
\usepackage{hyperref} 
\hypersetup{%
    colorlinks=true,
    linkcolor=blue,
    urlcolor=magenta
}
\usepackage[english]{babel}
\usepackage{subcaption}
\usepackage{amssymb}
\usepackage{bm}
\usepackage{booktabs}
\usepackage{graphicx}
\usepackage{rotating}
\usepackage[table]{xcolor} 

\usepackage{doi}

\definecolor{light-gray}{gray}{0.95}

\newcommand{\eps}{\varepsilon}
\renewcommand{\d}{\mathrm{d}}
\renewcommand{\vec}[1]{{\bm{#1}}}

\newcommand{\dA}{\,\mathrm{dA}}
\newcommand{\dV}{\,\mathrm{dV}}

\newcommand{\ExB}{$\bm{E}\times\bm{B}$} 

\newcommand{\RA}[1]{\left \langle #1 \right \rangle}

\newcommand{\sumsp}{\sum_{\mathrm{s}} }

\newcommand{\bhat}{\bm{\hat{b}}}
\newcommand{\bperp}{{ \vec b_\perp}}

\newcommand{\ehat}{\bm{\hat{e}}}

\renewcommand{\nperp}{\vec\nabla_{\perp}}
\newcommand{\nc}{\vec\nabla\cdot}
\newcommand{\cn}{\cdot\vec\nabla}
\newcommand{\vn}{\vec{\nabla}}
\newcommand{\npar}{\nabla_\parallel}

\newcommand{\KB}{ {\vec K_{\vn B}}}
\newcommand{\KK}{ {\vec K_{\vn\times\bhat}}}

\newcommand{\feltor}{FELTOR} 

\begin{document}
\title{%
Effects of plasma resistivity 
in three-dimensional full-F gyro-fluid turbulence simulations
%
}

\author{M. Wiesenberger$^1$ and M. Held$^{2,3}$}
\address{$^1$ Department of Physics, Technical University of Denmark, DK-2800 Kgs. Lyngby, Denmark}
\address{$^2$ Institute for Ion Physics and Applied Physics, Universit\"at Innsbruck, A-6020 Innsbruck, Austria}
\address{$^3$ Department of Mathematics and Statistics, UiT The Arctic University of Norway, 9037 Tromsø, Norway}
\ead{mattwi@fysik.dtu.dk}
\begin{abstract}

A full-F, isothermal, electromagnetic, gyro-fluid model is used to simulate plasma turbulence in a COMPASS-sized, diverted tokamak. 
A parameter scan covering three orders of magnitude of plasma resistivity
and two values for the ion to electron temperature ratio with otherwise fixed parameters is setup and analysed. Simulations are performed with a new version of the \feltor{} code, which is fully parallelized on GPUs. Each simulation covers a couple of milliseconds.

Two transport regimes for high and low plasma resistivities are revealed. Beyond a critical resistivity
the mass and energy confinement reduces with increasing resistivity. Further, for high plasma resistivity the direction of parallel acceleration is swapped compared to low resistivity. 

The integration of exact conservation laws over the closed field line region allows for an identification of numerical errors within the simulations. The electron force balance and energy conservation show relative errors on the order of $10^{-3}$ while the particle conservation and ion momentum balance show errors on the order of $10^{-2}$. 
 Relative fluctuations amplitudes increase from below $1\%$ in the core to $15\%$ in the edge and up to $40\%$ in the scrape-off layer.

Finally, three-dimensional visualisations using ray tracing techniques are displayed and discussed. The field-alignment of turbulent fluctuations in density and parallel current becomes evident.
\end{abstract}


\submitto{Plasma physics and controlled fusion}

\ioptwocol


\section{Introduction}

Turbulence in the edge and scrape-off layer (SOL) regions of magnetically confined plasmas displays very efficient (and unwelcome) transport properties~\cite{Wesson,ScottBook}. In fact, the observed levels of transport of particles and thermal energy out of the confined region by far exceed the ones predicted by collisional transport theory~\cite{Liewer1985,Wootton1990} even if neoclassical effects from the magnetic field geometry are taken into account. This has led to the alternative denomination of turbulent transport as "anomalous" transport.
Since particle and energy confinement are the ultimate goal of any magnetic fusion device plasma turbulence is subject to intensive research. 

Numerous challenges exist when modelling plasma turbulence.
For example, it is observed that relative fluctuation levels increase from the edge into the SOL and may approach and even exceed order
unity~\cite{McKee2001,Garcia2007,Zweben2007,Garcia2013,Zweben2015}. This was recently also found close to the X-point region~\cite{Nem2021}. 
This means that a linearisation of equations around a background profile
is inadmissible in modelling. Avoiding such a separation between stationary profile and dynamic fluctuations in models has the additional advantage that a profile can interact with turbulence and evolve self-consistently in time. The profile is then an output of the model rather than a given input.

Furthermore, it is observed that the ratio of ion-temperature relative to electron temperature is above one in the edge and scrape-off layer regions~\cite{Allan2016,Elmore2012,Kocan2012}. Turbulent eddies in the edge and
in blobs in the scrape-off layer are of the size $\rho_s = \sqrt{T_e m_i}/(eB_0)$ where $T_e$ and $m_i$ are electron temperature and ion mass respectively, $e$ is unit charge and $B_0$ is the reference magnetic field strength. With $\rho_i = \sqrt{T_im_i}/(eB_0) \approx \rho_s$ (with $T_i$ the ion temperature) this leads to finite Larmor radius and polarization effects being important for the dynamics of turbulent eddies and blobs~\cite{Wiesenberger2014blobs,Held2016blobs,Held2023NOBblobs}.

Full-F gyro-fluid models are able to evolve large fluctuation amplitudes, steep background profiles and include finite Larmor radius effects~\cite{Madsen2013,Wiesenberger2014blobs,Held2018NOB,Held2020Pade,Held2023NOBblobs}. Gyro-fluid models in general result from taking velocity space moments
over an underlying gyro-kinetic model and share many of its advantages: finite Larmor radius corrections, consistent particle drifts, an energy and momentum theorem based on variational methods in the underlying gyro-kinetic model and an inherent symmetry in moment equations with regards to multiple ion species. These advantages are absent from so-called drift-fluid models
that result from a drift-expansion of the Braginskii equations~\cite{Braginskii1965,Simakov2003,Madsen2016,Gath2019,Poulsen2020}. A downside of
gyro-fluid models, inherited again from their underlying gyro-kinetic models, are the impractical 
expressions for plasma-neutral interactions and scattering collisions available today. Attempts at numerically implementable expressions derived in a long-wavelength limit were recently presented in~\cite{Wiesenberger2022LWL}. 
Compared to gyro-kinetic models, gyro-fluid models invoke a closure scheme that can be tailored to specific physical regimes of interest, e.g. the collisional regime. Such closures can be adopted at the chosen number of moments, which emerge typically from a Hermite-Laguerre expansion in velocity space of the gyro-averaged gyro-center distribution function~\cite{Madsen2013,Held2020Pade}.
The number of moment equations is usually small (2 in the present work) and the associated reduced velocity space resolution translates to a corresponding saving in computational cost over gyro-kinetic models. This implies that gyro-fluid models are more computationally efficient for parameter scans or for resolving larger plasma volumes than gyro-kinetic models. 

Further challenges arise in numerical approaches to plasma turbulence.
The dynamics of a magnetized plasma is highly anisotropic
with respect to $\bhat$, the magnetic unit vector. Fluctuations along $\bhat$ typically
have a much larger extension $L_\parallel$ than fluctuations perpendicular to it $L_\perp \ll L_\parallel$. In a numerical simulation
 the use of field-aligned coordinates, in particular
flux-tube coordinate systems thus seems appropriate. The field alignment translates to a low spatial resolution requirement along the field line following coordinate \cite{Dewar1983,Beer1995,Scott2001}.
However, field aligned coordinate systems cannot include X-points in the geometry. 
This is a major downside as one or more X-points in the magnetic field equilibrium are a crucial ingredient to current tokamak design and in particular ITER~\cite{Aymar2002}. The X-point is connected
to the construction of a divertor, which separates the plasma-wall interactions
from the confined plasma region~\cite{Wesson}. Further, it plays a crucial role in and at least facilitates the transition to the high confinement mode~\cite{Wagner1985, Keilhacker1987, Connor2000}.
Correct modelling of magnetic field equilibria that include one or even several X-points is thus critical.

Two solutions to the problem exist to date. With the increase in computational resources it is possible to directly discretize and simulate model equations on non field-aligned coordinate systems~\cite{Halpern2016,Dorr2018}. This allows simulations including X-points as exemplified by the GBS~\cite{Giacomin2022}, STORM~\cite{Riva2019} or TOKAM-3X~\cite{Tamain2016} codes. However, such an approach does not
exploit the field-aligned character of turbulence and can thus only be used for small to medium sized tokamaks due to both strong numerical diffusion and extreme computational cost~\cite{Held2016FCI,Stegmeir2023}.
An alternative approach is the so-called flux-coordinate independent approach~\cite{Hariri2013,Held2016FCI,Stegmeir2017,Wiesenberger2023FCI}. Here, the grid is not field-aligned while at the same time the toroidal direction is resolved by only a few points. Turbulence simulations of AUG were successfully performed with the GRILLIX code~\cite{Stegmeir2019,Zholobenko2021}.

For the verification of codes the method of manufactured solution is often used~\cite{Dudson2016,Tamain2016,Halpern2016,Giacomin2022}.
However, even in two-dimensional turbulence simulations numerical errors on the order of machine precision exponentially increase to order one within a short period of time~\cite{Wiesenberger2019Feltor}. This is fundamentally due to the turbulent nature of the underlying model and not an issue of the numerical implementation.
Thus, turbulence simulations due to their very nature cannot reach pointwise convergence after sufficiently long simulation time. This
makes the method of manufactured solutions unsuitable for a verification of results on a long time span.

In this contribution we address the above challenges in a new version of the simulation code \feltor{}~\cite{Feltor,Wiesenberger2019Feltor}. 
As opposed to the drift-fluid models discretized in the mentioned GRILLIX, TOKAM-3X, GBS and STORM codes \feltor{}  discretizes
a full-F gyro-fluid model and thus benefits from finite Larmor radius effects, an exact energy conservation law and consistent particle drifts. 
Polarization effects are taken in the long wavelength limit in order to avoid inversion of an operator function~\cite{Held2023NOBblobs}.
Similar to the GRILLIX code \feltor{} uses an FCI scheme for its parallel dynamics but in a recently upgraded finite volume FCI formulation~\cite{Wiesenberger2023FCI} that has significantly improved conservation properties compared to previous versions~\cite{Hariri2013,Held2016FCI,Stegmeir2017}. For
the perpendicular dynamics \feltor{} chooses discontinuous Galerkin methods~\cite{NodalDG,Cockburn2001runge} in contrast to the above codes, which rely on finite difference methods. \feltor{} is the only code among the mentioned ones that is fully ported to GPUs using a platform independent MPI+X implementation.
Recently, all the above codes including \feltor{} were part of a validation effort in TORPEX and TCV~\cite{Galassi2022,Oliveira2022}.

 \feltor{} allows stable simulations encompassing several milliseconds of turbulent dynamics. The simulations are computationally cheap enough that a parameter scan is possible. We vary the plasma resistivity and the ion to electron temperature ratio in $12$ simulations.
 We present techniques for three-dimensional visualisations using ray-tracing in order to gain visual intuition of the magnetic field, the density and the parallel current. In particular the field-alignment of turbulent fluctuations with $L_\parallel \ll L_\perp$  is visible.
 In order to quantitatively analyse the simulation data we introduce the flux-surface averages and integration. Numerically, these are accurately computed by transforming on a flux-aligned grid~\cite{Wiesenberger2023FCI}. We discuss flux-surface averaged density and fluctuation profiles.
Afterwards, we focus on verification of the implementation. Since, as pointed out above, pointwise long-term convergence tests are impossible we here present verification through exact analytical conservation laws. These include mass, parallel momentum and energy conservation as well as the electron force balance. We suggest to use volume and time integration to define a numerical error of simulation results. At the same time we are able to identify the largest and most important terms in each of the mentioned conservation equations and further in the total parallel momentum balance.
Applied to the mass and energy conservation, we can compute and discuss the mass and energy confinement times. The latter relate back to our initial statement of confinement being an important
goal for the magnetic fusion devices.

This work is structured as follows. In
Section~\ref{sec:model} we present the gyro-fluid model including resistivity and diffusive terms, the density source and boundary conditions. This is followed by Section~\ref{sec:magnetic} where the magnetic field is described. A parameter scan over plasma resistivity and ion temperature is setup for model simulations of the COMPASS tokamak in Section~\ref{sec:setup} discussing the COMPASS magnetic field and the exact physical parameters in use.
In Section~\ref{sec:results} we present the results. We discuss performance observations, three-dimensional visualisations and density and fluctuation profiles. In particular, here we show the numerical verification with a focus on mass, energy, ion momentum and parallel force balance. Finally, we discuss particle and energy confinement times computed from previously analysed terms in the mass and energy conservation equations. We conclude in Section~\ref{sec:conclusion}.



\section{The model} \label{sec:model}
In the following we denote $\phi$ as the electric potential, $A_\parallel$ the parallel magnetic potential, $m$ the species mass, $q$ the species charge, $N$ the gyro-centre density, $U_\parallel$ the gyro-centre parallel velocity, $T_\perp$, $T_\parallel$ the perpendicular, parallel temperatures, $\bhat$ the magnetic unit vector field and $B$ the magnetic field strength.
Note that all species dependent quantities $m$, $q$, $N$, $U_\parallel$, $T_\perp$ and $T_\parallel$ have an implied species index $s$ that we omit in the notation.
We define two magnetic field curvature vectors
\begin{align}
    \KK :=& \frac{1}{B}(\vn \times \bhat), \\
    \KB :=& \frac{1}{B}(\bhat \times\vn \ln B),
\end{align}
as well as perpendicular and parallel derivatives
\begin{align}
    \nperp :=& -\bhat\times ( \bhat \times \vn), \qquad
&\Delta_\perp :=&\nc \nperp, \\
\npar :=& \bhat \cn, \qquad
&\Delta_\parallel :=& \nc \bhat \bhat \cn .
\end{align}
Notice the formulary in~\ref{sec:formulary}.

\subsection{Gyro-fluid moment equations}
The gyro-centre continuity and parallel momentum conservation equations read for each species~\cite{Madsen2013,Held2020Pade,WiesenbergerPhD,HeldPhD} (omitting the species label)
\begin{align}\label{eq:density}
    \frac{\partial}{\partial t} N +& \nc \vec J_N = \Lambda_N + S_N, \\
\frac{\partial}{\partial t} \left(mN U_\parallel\right) +& qN \frac{\partial}{\partial t} A_\parallel 
+ \nc \vec J_{mNU} \nonumber \\
=& F_{mNU,\vn B} + F_{mNU,\psi}  + R_\parallel +  \Lambda_{mNU}.
\label{eq:parallel-mom}
\end{align}
The system is closed by the parallel Ampere law 
\begin{align}\label{eq:ampere}
    -\mu_0 \Delta_\perp A_\parallel = \sumsp q NU_\parallel
\end{align}
and the polarisation equation
\begin{align}  \label{eq:polarisation}
   \sumsp \left[ q\Gamma_1  N + \nc\left( \frac{mN}{B^2} \nperp \phi\right)\right] = 0,
\end{align}
where we sum over all species. 
We have the density current
\begin{align}   \label{eq:density-flux}
\vec J_N :=&  NU_\parallel(\bhat + \bperp) 
 +N \frac{\bhat \times\vn  \psi}{B}
    \nonumber\\
    &+\frac{NT_\parallel + mNU_\parallel^2}{q} 
        \KK + \frac{NT_\perp}{q} \KB ,
\end{align}
momentum current
\begin{align}
\vec J_{mNU} :=&  (mNU_\parallel^2+NT_\parallel) (\bhat + \bperp) \nonumber\\
&+mN U_\parallel\frac{\bhat\times  \vn\psi }{B} 
 \nonumber\\
    & +m\frac{3U_\parallel NT_\parallel + mNU_\parallel^3}{q} \vec \KK \nonumber\\
    &+m\frac{U_\parallel NT_\perp}{q} \KB
    \label{eq:parallel-mom-current}
\end{align}
and the electric and mirror force terms
\begin{align}
    F_{mNU,\psi} =& -qN(\bhat +
    \bperp) \cdot \vn\psi \nonumber\\
    &-m N U_\parallel\KK\cdot\vn \psi ,
    \label{eq:parallel-electric}
\\
    F_{mNU,\vn B} =& -NT_\perp (\bhat + \bperp)\cn \ln B 
    \nonumber\\
    &-m\frac{U_\parallel NT_\perp}{q} \KK\cn\ln B. \label{eq:parallel-mirror}
\end{align}
 The definition of the diffusive terms $\Lambda_N$ and $\Lambda_{mNU}$ and the resistivity $R_\parallel$ are shown in Section~\ref{sec:dissres} while the gyro-centre density source term $S_N$ is defined in Section~\ref{sec:sources}. No source is added in the parallel momentum equation.
We use
\begin{align}
    \Gamma_1 :=& \left(1-\frac{\rho_0^2}{2} \Delta_\perp\right)^{-1},\qquad
    \rho_0^2 := \frac{mT_\perp}{q^2B_0^2}, \\
    \bperp :=& \frac{\vn\times A_\parallel\bhat}{B} = A_\parallel \KK + \frac{\vn A_\parallel \times\bhat}{B},\\
    \psi :=& \Gamma_1(\phi) -\frac{m}{2qB^2} |\nperp \phi|^2 ,\\
    T_\perp = & T_\parallel = T = const.
\end{align}
These are the Pad{\'e} approximated gyro-average operator $\Gamma_1$ with thermal gyro-radius $\rho_0$, the perpendicular magnetic field perturbation $\bperp$, the gyro-centre potential $\psi$ and temperature $T$.

We keep a 2nd order accurate gyro-averaging operator $\Gamma_1$ independent of particle position that closely mimics an exponential to arbitrary order~\cite{Held2020Pade}. The polarisation in the second term in Eq.~\eqref{eq:polarisation} is taken in a long wavelength limit while all finite Larmor radius effects are neglected in the parallel magnetic potential $A_\parallel$.

In Eq.~\eqref{eq:density-flux} we can identify the density flux parallel to the magnetic field $\bhat$ perturbed by magnetic fluctuations $\bperp$, followed by the \ExB{}, the curvature and the grad-B drifts. 

The first term in the momentum current Eq.~\eqref{eq:parallel-mom-current} consists of the parallel momentum current quadratic in the parallel velocity $U_\parallel$. This term is an expression of Burger's term and can lead to shocks if no parallel viscosity was added to the system. 
The term $\nc(NT_\parallel (\bhat + \bperp))$ stemming from $\nc \vec J_{mNU}$ with $\vec J_{mNU}$  from Eq.~\eqref{eq:parallel-mom-current} can be combined with the mirror force $NT_\perp(\bhat+\bperp)\cn \ln B$ in Eq.~\eqref{eq:parallel-mirror} to yield the familiar pressure gradient $(\bhat + \bperp) \cn (NT)$ with the identity $\nc ( \bhat + \bperp) = -(\bhat + \bperp) \cn \ln B $  and the assumption $T_\perp = T_\parallel = T$. Further, in Eq.~\eqref{eq:parallel-mom-current} we have the \ExB{} and curvature drift transport of parallel momentum. In the parallel electric force Eq.~\eqref{eq:parallel-electric} we have the parallel and perturbed gradients of the gyro-centre electric potential $\psi$ together with a correction due to the magnetic curvature. Even though the latter term is small it must be kept to guarantee energetic consistency. The equivalent correction also appears in the mirror force term Eq.~\eqref{eq:parallel-mirror}.

\subsection{Simplifications}
\subsubsection{Two species}
Even though the model is formulated inherently as a multi-species model we here only treat an electron-ion plasma, specifically with Deuteron ions ($q_i=e$, $m_i\approx 2m_p$ with $m_p$ the proton mass). The model can also be used to simulate electron-positron plasmas~\cite{Kendl2017}. Multi-species gyro-fluid simulations were presented in~\cite{Meyer2016,Reiter2023}. 
\subsubsection{Small electron mass}
We take the electron gyro-radius to be zero $\rho_{0,e} = 0$ and thus have~\cite{Wiesenberger2014blobs,Held2016blobs}  
\begin{align}
\Gamma_{1,e} = 1, \qquad
\psi_e = \phi.
\end{align}
This is combined with neglecting the electron mass in the polarisation equation, which thus reads
\begin{align} \label{eq:polarisation-simplified}
-en_e + q\Gamma_{1,i} N_i + \nc \left( \frac{m_iN_i}{B^2} \nperp \phi\right) = 0.
\end{align}
Note here that we denote the electron gyro-centre density as $n_e$ and gyro-centre parallel velocity as $u_{\parallel,e}$ (as opposed to $N_e$ and $U_{\parallel,e}$) to signify that these quantities coincide with the actual fluid particle density and parallel particle velocity.

\subsubsection{Toroidal field line approximation} \label{sec:torfieldlineapprox}
The toroidal field line approximation applies \(\bhat= \pm \ehat_\varphi\) to all perpendicular operators
(e.g.: perpendicular elliptic operator and curvature operators)
but retains the full expression for the magnetic field unit vector \(\bhat\)
in parallel operators \(\npar\) and \(\Delta_\parallel \)~\cite{WiesenbergerPhD,HeldPhD}.
Note that we allow the negative sign $-\ehat_\varphi$ to enable a sign reversal of the magnetic field.

We employ cylindrical coordinates \( (R,Z,\varphi) \), with \(\varphi\) anti directed to the geometric toroidal angle ({\bf clockwise} if viewed from above) to
obtain a right handed system.
This yields\begin{align}
\bhat \times \vn f \cn g &\approx\pm\ehat_\varphi \times \vn f  \cn g = \pm\ehat_\varphi \cdot (\vn f \times \vn g)
\nonumber\\
&= \pm\frac{1}{R} \left(\frac{\partial f}{\partial R}\frac{\partial g}{\partial Z} - \frac{\partial f}{\partial Z}\frac{\partial g}{\partial R}\right), \\
\nperp f &\approx \frac{\partial f}{\partial R} \ehat_R + \frac{\partial f}{\partial Z}\ehat_Z, \\
\Delta_\perp f &\approx \frac{1}{R}\frac{\partial}{\partial R} \left( R \frac{\partial f}{\partial R}\right) + \frac{\partial }{\partial Z}\left(\frac{\partial f}{\partial Z} \right).
\label{}
\end{align}
The curl of $\bhat$ reduces to
 $\vn\times\bhat \approx -  \frac{\pm 1}{R} \ehat_Z$.
This simplifies the curvature operators to:
\begin{align} 
\KK  &\approx  -  \frac{\pm 1}{B R} \ehat_Z , \nonumber\\
\KB  &\approx  -\frac{\pm 1}{B^2}\frac{\partial B}{\partial Z}\ehat_R +\frac{\pm 1}{B^2} \frac{\partial B}{\partial R}\ehat_Z & \label{eq:curvature-approx}
\end{align}
and
\begin{align}
    \nc \KK &= \frac{\pm 1}{R B^2} \frac{\partial B}{\partial Z} = -\nc \KB,
\end{align}
which maintains a vanishing divergence of the total curvature \( \nc \vec{ {K} } = 0\) with $\vec K := \KK+\KB$.

The toroidal field approximation is motivated numerically. The true perpendicular derivatives contain derivatives in the $\varphi$ direction, which would have to be resolved numerically. Since we expect turbulent eddies to be highly elongated along the field lines but very narrow perpendicular to $\bhat$ this translates to a very high resolution requirement in the $\varphi$ direction. The toroidal field approximation in combination with the FCI approach avoids this.

\subsection{Resistivity and diffusive terms}\label{sec:dissres}
Here, we discuss the terms $\Lambda_N$ in Eq.~\eqref{eq:density} and $\Lambda_{mNU}$, $R_\parallel$ in Eq.~\eqref{eq:parallel-mom}.
These terms take the form
\begin{align}
\Lambda_{N} :=&  -\mu_{N,\perp}(-\Delta_\perp)^2 N + \mu_{N,\parallel}\Delta_\parallel N \equiv -\nc \vec j_{N,\nu},\label{eq:perpdiffN}
\end{align}
with $\vec j_{N,\nu} := - \mu_{N,\perp} \nperp (-\Delta_\perp N) - \mu_{N,\parallel} \bhat \npar N$,
\begin{align}
\Lambda_{m_en_eu_e}  :=&   -\mu_{U,\perp}(-\Delta_\perp)^2 u_{\parallel,e}  + \mu_{\parallel,e} \Delta_\parallel u_{\parallel,e} 
\nonumber\\&-\nc ( m_e u_{\parallel,e} \vec j_{n_e,\nu}) ,
\nonumber\\
\Lambda_{m_iN_iU_i} :=&  -\mu_{U,\perp}(-\Delta_\perp)^2 U_{\parallel,i} + \mu_{\parallel,i} \Delta_\parallel U_{\parallel,i}   
\nonumber\\&-\nc ( m_i U_{\parallel,i} \vec j_{N_i,\nu}), \label{eq:perpdiffU}
\end{align}
and
\begin{align}
R_\parallel:=& -\eta_\parallel eq n_e ( N_i U_{\parallel,i} - n_e u_{\parallel,e}). \label{eq:parallel-resistivity}
\end{align}
We first notice that the diffusion terms have the form of total divergences $\Lambda_N = -\nc j_{N,\nu}$ and $\Lambda_{mNU} =: -\nc (\vec { \bar j}_{mNU,\nu} + mU_\parallel \vec j_{N,\nu} )$. Under volume integration these terms vanish modulo surface terms, which is important for mass and momentum conservation. 
Second, we notice the term $-\nc ( m U \vec j_{\nu,N})$
in the momentum diffusion~\eqref{eq:perpdiffU} has the form of a velocity convection. This is a correction term
that prevents energy from being generated by mass diffusion
as we will see explicitly in Section~\ref{sec:energy} and was suggested by for example \cite{Guermond2016,Wiesenberger2023FCI}.

The consistent treatment of the diffusive terms is particularly important for the parallel ion momentum equation. The alternative variant $\Lambda_{mNU,\parallel}:=\mu_{\parallel}\Delta_\parallel U_\parallel +\mu_{N,\parallel} m U_\parallel \Delta_\parallel N$ has the advantage that in velocity formulation $\Lambda_{U,\parallel} = \mu_{\parallel}\Delta_\parallel U_\parallel / (m N)$ simplifies~\cite{Stegmeir2019}. However, in this formulation the term $\mu_{N,\parallel} m U_\parallel \Delta_\parallel N$
unphysically generates momentum, leading to artificial toroidal rotation after a long enough simulation time. Other works on drift-fluid models completely neglect the parallel ion and electron viscosities~\cite{Tamain2016,Riva2019,Giacomin2022}.

In Eqs.~\eqref{eq:perpdiffN} and \eqref{eq:perpdiffU}, $\mu_{N,\perp}$ and $\mu_{U,\perp}$ are ad-hoc artificial numerical diffusion coefficients that are added to stabilize perpendicular advection and are thought to be small.
In the same sense $\mu_{N,\parallel}$ represents artificial parallel diffusion necessary to stabilize the parallel advection~\cite{Wiesenberger2023FCI}. 

The parallel velocity difference \( u_{\parallel,i} - u_{\parallel,e}:=(N_iU_{\parallel,i} - n_eu_{\parallel,e})/n_e\)
determines the parallel resistive term $R_\parallel$ in Eq.~\eqref{eq:parallel-resistivity}.
The term is positive for electrons with $q_e = -e$ and negative for ions with $q_i = e$.
This form both conserves parallel momentum and vanishes for zero current but
leads to a quadratic energy dissipation term only in the long-wavelength limit as we discuss in Section~\ref{sec:energy}. 

For the parallel viscosity $\mu_\parallel$ and the parallel resistivity $\eta$ we
copy the parallel resistive and viscous terms from the Braginskii fluid equations~\cite{Braginskii1965}.
The electron-ion and ion-ion collision frequencies are given by
$\nu_{ei} = \sqrt{2} z^2 e^4 \ln \Lambda n_e / (12\pi^{3/2} \sqrt{m_e} \epsilon_0^2 T_e^{3/2})$, $\nu_{ee} = \nu_{ei}/\sqrt{2}$
and
$\nu_{ii} =  z^4 e^4 \ln \Lambda n_i / (12\pi^{3/2} \sqrt{m_i} \epsilon_0^2 T_i^{3/2}) = \nu_{ei} \sqrt{m_e/m_i}/ ( (T_i/T_e)^{3/2} \sqrt{2})$.
We define with the parallel Spitzer resistivity
$\eta_\parallel := 0.51\frac{ m_e \nu_{ei}}{n_e e^2}$ and the parallel electron and ion viscosities
$\mu_{\parallel,e}:=0.73\frac{n_eT_e}{\nu_{ei}}$ and $\mu_{\parallel,i} = 0.96\frac{n_iT_i}{\nu_{ii}}$~\cite{Braginskii1965} the dimensionless parameter
\begin{align}
    \eta&:=\frac{en_0\eta_\parallel}{B_0} = 0.51\frac{\nu_{ei,0}}{\Omega_{e0}}\nonumber\\
    &=
    8.45\cdot 10^{-5}\ln \lambda \left(\frac{n_0}{10^{19}\text{m}^3}\right)
    \left(\frac{T_e}{\text{eV}}\right)^{-3/2}
    \left(\frac{B_0}{\text{T}}\right)^{-1},
    \label{eq:resistivity}
\end{align}
with $\nu_{ei,0} := \nu_{ei}(n_0, T_e)$
as well as
\begin{align}
    \nu_{\parallel,e}&:=\frac{\mu_{\parallel,e}}{m_e n_0\rho_s^2\Omega_{i0}}
    = 0.73 \frac{\Omega_{e0}}{\nu_{ei,0}} = \frac{0.37}{\eta},
    \label{eq:nu_parallele}\\
    \nu_{\parallel,i}&:=\frac{\mu_{\parallel,i}}{m_i n_0 \rho_s^2\Omega_{i0}}
    = 0.96 \frac{\Omega_{0}}{\nu_{ii,0}} = {\left(\frac{T_i}{T_e}\right)^{3/2}}{\sqrt{\frac{m_e}{m_i}}}
    \frac{0.69}{\eta},
    \label{eq:nu_paralleli}
\end{align}
with $\ln \lambda \approx 10$, $\Omega_{i0} = eB_0/m_i$ the ion gyro-frequency and $\Omega_{e0} = eB_0/m_e$ the electron gyro-frequency.
Finally, in order to prevent unreasonably small simulation timestep we need to impose a maximum and minimum on $\nu_{\parallel,e}$ and $\nu_{\parallel,i}$:
\begin{subequations} \label{eq:restriction}
\begin{align}
    \nu_{\parallel,e} = \min &{ \left(\frac{0.37}{\eta},\ \frac{0.37}{10^{-4}} \right)}, \\
    \nu_{\parallel,i} = \min&\left( \max{\left( \sqrt{\frac{m_e}{m_i}}\frac{0.69}{10^{-4}},\ \left(\frac{T_i}{T_e}\right)^{3/2}\sqrt{\frac{m_e}{m_i}}\frac{0.69}{\eta} \right)},\right.
    \nonumber\\& \left. \frac{0.37}{10^{-4}}\right).
\end{align}
\end{subequations}
We emphasize that this restriction is numerically motivated.
The physical implications of Eq.~\eqref{eq:restriction} are discussed in Section~\ref{sec:results}. 
\subsection{Sources} \label{sec:sources}
We provide a constant influx of particles
\begin{align} \label{eq:source}
    S_{n_e}(R,Z,\varphi, t) &= \omega_s n_\mathrm{s}(R,Z),
\end{align}
where $\omega_s$ is the source strength parameter and $n_\mathrm{s}(R,Z)$ is an in principle arbitrary toroidally symmetric profile, which we discuss further in Section~\ref{sec:profiles}.  
In order to not generate potential with the source term the
ion gyro-centre source needs to fulfill $S_{n_e} = \Gamma_{1,i}S_{N_i} + \nc\left( \frac{m_i S_{N_i}}{B^2}\nperp \phi\right)$ for given particle source $S_{n_e}$ and potential $\phi$, which follows from a time derivative of Eq.~\eqref{eq:polarisation}. We were unable to invert this equation numerically. Only in the long wavelength limit can it be inverted to yield the approximation~\cite{Wiesenberger2022LWL}
\begin{align}
    S_{N_i} \approx \left(1-\frac{1}{2}\rho_{0i}^2 \Delta_\perp\right) S_{n_e} -\nc\left( \frac{m_i S_{n_e}}{B^2}\nperp \phi\right).
  \label{eq:ion_source}
\end{align}
The long wavelength limit should be well-fulfilled for a realistic source term since the amplitude $\omega_s$ is typically quite small.
Note that the additional terms besides $S_{n_e}$ in Eq.~\eqref{eq:ion_source} are total divergences, which means
they do not change the volume integrated "total" particle number created by the source.

A second task of the source $S_N$ is to globally ensure a minimum density.
This is required since through sheath dissipation the density can in principle become arbitrarily close to zero. 
This is, however, both detrimental to the stability of the simulation as well as the CFL condition (and thus the allowed time step) of the simulation and in reality also never happens due to e.g. wall-recycling. For both electrons and ions we choose the additional source term
\begin{align}  \label{eq:minimumne}
    S_{N,\min} = -\omega_{\min} ( N - n_{\min} ) H_{\alpha/2} ( n_{\min} - \alpha/2 - N),
\end{align}
where $H_\alpha(x)$ is a continuously differentiable approximation to the Heaviside function with width $\alpha$.
The Heaviside function ensures that this source term only acts when the density is below the lower limit. In our simulations
we choose $\omega_{\min} = 1$, $n_{\min} =0.2 n_0$, $\alpha = 0.05$.
\subsection{Boundary conditions} \label{sec:boundary}
Following~\cite{Stegmeir2019} we setup boundary conditions with the immersed boundary method using volume penalization~\cite{Schneider2015}.
In order to do this we first formally define a wall function 
\begin{align}
\chi_w( \vec x) = \begin{cases} 
1 \text{ for } \vec x \in \Omega_w \\
0 \text{ else}
\end{cases},
\end{align}
where $\Omega_w$ is the wall domain. Analogously, a sheath function $\chi_s$ can 
be defined using a sheath domain $\Omega_s$. Both $\chi_w$ and $\chi_s$ are further specified in Section~\ref{sec:flux}. We have $\Omega_s \cap \Omega_w = \emptyset$. We can then enforce boundary conditions on the wall and sheath by
\begin{subequations} \label{eq:sheath_wall_equations}
    \begin{align}
        \frac{\partial}{\partial t} N =& F_N ( 1-\chi_s - \chi_w)  -\omega_s\chi_s (N-N_{sh}) 
        \nonumber\\
        &-\omega_w\chi_w (N-N_{w}),\\
       \frac{\partial}{\partial t} (mU_\parallel + qA_\parallel)=& \frac{mF_{mNU} - mU_\parallel F_N}{N}  ( 1 - \chi_s- \chi_w)
       \nonumber\\
       &-m\omega_s \chi_s (U_\parallel - U_{\parallel,sh})
        \nonumber\\
        &-m\omega_w\chi_w (U_\parallel- U_{\parallel,w}),
    \end{align}
\end{subequations}
where $F_N:= - \nc \vec j_N - \nc \vec J_N + \Lambda_N + S_N$ follows from Eq.~\eqref{eq:density} and
 $F_{mNU} = - \nc \vec J_{mNU} + F_{mNU,\vn B} + F_{mNU,\psi}  + R_\parallel +  \Lambda_{mNU}$ follows from Eq.~\eqref{eq:parallel-mom}.
We choose $\omega_s = 5$ and $\omega_w = 0.01$.
The polarization equation is penalized according to the immersed boundary method
\begin{align}\label{eq:polarization_penalized}
    -\nc\left( \frac{N_i}{B^2}\nperp \phi \right) = (\Gamma_{1,i} N_i - n_e)(1-\chi_w -\chi_s).
\end{align}
We do not penalize the parallel Ampere law due to numerical stability.

We choose the wall conditions $N_w = 0.2$ and $U_{\parallel,w} = 0$. Further, we have $\phi_w = 0$ and $\nperp A_{\parallel,w} = 0$ for the electric and magnetic potential. The latter two are however only enforced at the domain boundaries rather than through a penalization method.
We have the insulating sheath boundary conditions
\begin{align}
    U_{\parallel,i, sh} &= \pm\sqrt{\frac{T_e+T_i}{m_i}}, \\
    u_{\parallel,e, sh} &= U_{\parallel,i,sh}N_i/n_e.
\label{eq:insulating_sheath} 
\end{align}
$N_{sh}$ is chosen such that $\npar N|_{sh} = 0$. 

\section{The magnetic field}\label{sec:magnetic}

This section discusses \feltor{}'s general capabilities to represent toroidally
symmetric magnetic fields. The specific magnetic field used for the main physical discussion in Section~\ref{sec:results} is presented in Section~\ref{sec:flux}.

\subsection{The flux function}
In cylindrical coordinates the general axisymmetric magnetic field obeying an MHD equilibrium ($\mu_0 \vec j = \vn\times \vec B$, $\vn p = \vec j \times \vec B$) can be written as~\cite{haeseleer}
\begin{align} \label{eq:magnetic-field}
 \vec{B} &= \frac{1}{R}\left[I(\psi_p) \ehat_{\varphi} + \frac{\partial
 \psi_p}{\partial Z} \ehat_R -  \frac{\partial \psi_p}{\partial R} \ehat_Z\right].
\end{align}
Here, $\psi_p$ is the poloidal flux function and $I(\psi_p)$ is the current stream function.
For the sake of clarity we define the poloidal magnetic field \( \vec{B}_p = \frac{1}{R}\left( \frac{\partial \psi_p}{\partial Z}\ehat_R - \frac{\partial \psi_p}{\partial R}\ehat_Z\right)
\) and the toroidal magnetic field \(\vec{B}_t =\frac{I}{R} \ehat_{\varphi}\).

Note that
with a typically convex function $\psi_p$ (second derivative is
positive), $I(\psi_p)>0$ and the previously defined coordinate system the field
line winding is a left handed screw in the positive $\ehat_\varphi$-direction.
Also note that then $\vec B\times\vn\vec B$ points down, which for a lower single null configuration is towards the magnetic X-point,
and we have the {\bf favourable} drift direction (in experiments H-mode
is reached more easily in this configuration~\cite{Snipes1996,Connor2000,Chen2018}).

We have the contravariant components of $\vec B$
\begin{align}
    B^R = \frac{1}{R} \frac{\partial \psi_p}{\partial Z}, \quad
    B^Z = - \frac{1}{R}  \frac{\partial \psi_p}{\partial R}, \quad
    B^\varphi  = \frac{I}{R^2}
\end{align}
and the covariant components $B_R = B^R$, $B_Z = B^Z$ and $B_\varphi = R^2B^\varphi$.
By construction we have $\partial_\varphi B = 0$ with
\begin{align}
  B = \frac{1}{R}\sqrt{ {I^2 + |\vn \psi_p|^2}}.
    \label{eq:magnetic-field-strength}
\end{align}
In \feltor{} we have various ways to represent the flux function $\psi_p$ and its derivatives.
In this work we use a general solution to the Grad-Shafranov equation using Solov'ev pressure and current profiles~\cite{Cerfon2010,Cerfon2014}
\begin{subequations}
\label{eq:solovev}
\begin{align}
 \psi_p (R,Z) &= \mathcal P_{\psi} B_0R_0^2 \left[ A\left( \frac{1}{2} \bar{R}^2 \ln{\bar{R}}
   - \frac{1}{8}\bar{R}^4\right) + \frac{1}{8}\bar{R}^4 \right .\nonumber\\ 
   &\left. + \sum_{i=1}^{12} c_{i}  \bar{\psi}_{pi}(\bar R, \bar Z )\right], \\
   I(\psi_p) &= \mathcal P_IB_0R_0\sqrt{ - 2A\frac{\psi_p}{\mathcal P_{\psi}B_0R_0^2} +1},
\end{align}
\end{subequations}
with $A$, $\mathcal P_\psi$ free constants, $\mathcal P_I = \pm \mathcal P_\psi$ for $A\neq 0$ and $\mathcal P_I$ arbitrary for $A=0$ (purely toroidal equilibrium current).
We introduce \(\bar{R} \equiv {R}/{R_0}\) and \(\bar{Z} \equiv{Z}/{R_0}\) where $R_0$ is the major radius and $B_0$ is a reference magnetic field strength.
The dimensionless base functions $\bar \psi_{pi}$ are listed in \cite{Cerfon2010}.


\subsection{Discussion}
Since Eqs.~\eqref{eq:solovev} is given in terms of analytical base functions we can numerically evaluate $\psi_p(R,Z)$ and $I(\psi_p)$
and all their derivatives
at arbitrary points to machine precision, which is simple to implement and fast to execute.
This translates to an exact representation of the magnetic field and related
quantities, for example curvature~\eqref{eq:curvature-approx}, in code. In particular,
the X-point(s) and O-point can be determined to machine
precision via a few Newton iterations.

The choice of the coefficients \(c_{i}\) and \(A\) determines the actual form
of the magnetic field.
We can for example represent single and asymmetric double X-point configurations, force-free states,
field reversed configurations and low and high beta tokamak equilibria~\cite{Cerfon2010,Cerfon2014}.
The scaling factors $\mathcal P_\psi$ and $\mathcal P_I$ are mainly introduced to maximize the flexibility e.g. to adapt the solution to experimental equilibria or to reverse the sign of the magnetic field.

If one or more X-points are present, we choose $c_1$ such that
$\psi_p(R_X, Z_X) = 0$ for the X-point closest to the O-point that is the separatrix is given by $\psi_p(R,Z) = 0$.

We offer several predefined sets of parameters as well as Mathematica and Python scripts to generate / fit coefficients to experimental equilibria in the \url{https://github.com/feltor-dev/magneticfielddb} repository. The contained Jupyter Notebooks and Python scripts help setting up appropriate simulation domains as well as wall and sheath regions $\chi_w$ and $\chi_s$ as presented in Section~\ref{sec:flux}. See~\ref{app:data} for  more details.

\section{Simulation setup} \label{sec:setup}

\subsection{The magnetic flux, the wall and the sheath} \label{sec:flux}
\begin{figure*}[htbp]
    \centering
\begin{subfigure} {0.45\textwidth}
    \includegraphics[width = \textwidth]{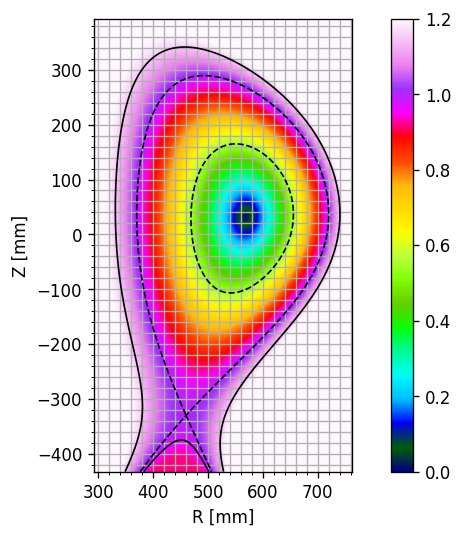}
    \caption{Normalized poloidal flux $\rho_p$ Eq.~\eqref{eq:rhop}} \label{fig:fluxa}
    \end{subfigure}
    \begin{subfigure} {0.45\textwidth}
    \includegraphics[width = 0.79\textwidth]{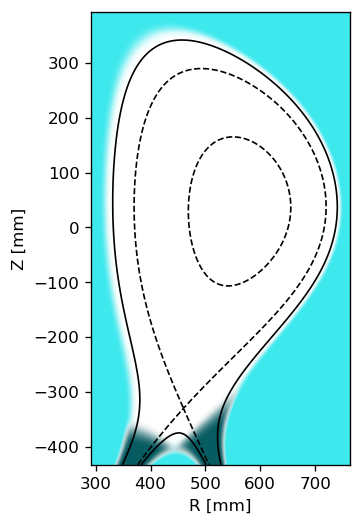}
    \caption{Wall $\chi_w$ and sheath $\chi_s$ function } \label{fig:fluxb}
    \end{subfigure}
    \caption{Calibration of the simulation box. The normalized magnetic flux $\rho_p = \sqrt{(\psi_{p,O} - \psi_p)/\psi_{p,O}}$ on the left and the wall and sheath regions on the right. The magnetic flux $\psi_p$ is modified to a constant inside the wall region. On the right plot colours range linearly from $0$ to $1$. Two contour lines indicating the wall at $\rho_p = 0.97$ in the private flux region and $\rho_p = 1.15$ in the scrape-off layer region are plotted in solid black lines. The separatrix $\rho_p=1$ and the boundary of the source region  at $\rho_{p} = 0.55$ in the core are plotted in black dashed lines.}
    \label{fig:flux}
\end{figure*}

The first step in setting up a simulation with \feltor{} is to choose an appropriate magnetic field. 
In this work we choose to model the COMPASS tokamak and fit the magnetic flux function described in \cite{Panek2016} with a Solov'ev equilibrium described in Eq.~\eqref{eq:solovev}. One X-point is
situated at $R_X=460$ mm, $Z_X=-330$ mm with  $\psi_p(R_X,Z_X)=0$ and the O-point is situated at $R_O=568.78$ mm, $Z_O=32.69$ mm with $\psi_{p,O} := \psi_p(R_O,Z_O) = -18.76\rho_s R_0 B_0$ (found with a few iterations of a Newton solver).
In Fig.~\ref{fig:fluxa} we plot the normalized poloidal flux
\begin{align}
\rho_p = \sqrt{\frac{\psi_{p,O} - \psi_p}{\psi_{p,O}}}.
\label{eq:rhop}
\end{align}
In Fig.~\ref{fig:fluxb} we plot the chosen wall and sheath functions $\chi_w$ and $\chi_s$, which signify the penalization regions for the immersed boundary conditions in Eq.~\eqref{eq:sheath_wall_equations} and Eq.~\eqref{eq:polarization_penalized}. The wall region is given simply as a flux aligned region
\begin{align}
    \chi_w(R,Z) = \begin{cases}
        1 \text{ if} &\rho_p(R,Z) > \rho_w \vee \\ 
        &( \rho_p(R,Z) < \rho_{F} \wedge Z < Z_X ) \\
        0 \text{ else}&
    \end{cases}.
\end{align}
Here we choose $\rho_w = 1.15$ for the scrape-off layer and the private flux region at $\rho_F = 0.97$.
For the sheath region we first define an angular distance $\varphi_w$ of each point $(R,Z)$ to the bounding box via the integration of
\begin{align}\label{eq:sheath_coordinate}
    \frac{dR}{d\varphi} = \frac{b^R}{b^\varphi}, \qquad
    \frac{dZ}{d\varphi} = \frac{b^Z}{b^\varphi},
\end{align}
with initial condition $(R,Z)$ until $R((\varphi_w), Z(\varphi_w))$ intersects the bounding box. The intersection can be found with a bisection algorithm. 
The sheath is then given by
\begin{align}
    \chi_s(R,Z) := \begin{cases} 
        1 \text{ if } \varphi_w(R,Z) > \varphi_0  \\
        0 \text{ else}
    \end{cases},
\end{align}
where we choose $\varphi_0 = 7/32$.
Note that for numerical reasons we implement a continuously differentiable transition at the boundary of
the regions $\Omega_w$ and $\Omega_s$. 

Both plots in Fig.~\ref{fig:flux} show the numerical simulation domain in the $R$-$Z$ region as $[R_0,R_1]\times [Z_0, Z_1]$.

\subsection{Initial profiles and sources} \label{sec:profiles}
To initialize our simulation we choose
\begin{align}\label{eq:density_profile}
N(R,Z,\varphi,0) =& n_{\text{prof}}(R,Z) \nonumber\\
:=&
        (n_{\text{peak}}-n_{\text{sep}})\frac{\psi_p(R,Z)}{\psi_{p,O}} + n_{\text{sep}},
\end{align}
equal for both electrons and ions such that the profile given in~\cite{Panek2016} is approximately reproduced with a peak density of $n_{\text{peak}} = 8.5\cdot 10^{19}$m$^{-3}$
and a separatrix density of $n_{\text{sep}} = 10^{19}$m$^{-3}$. In the SOL
the profile exponentially decreases to the background density of $n_{\min} = 0.2\cdot 10^{19}$m$^{-3}$.

The initial parallel velocity for both electrons and ions is zero everywhere except in the scrape-off layer where it varies linearly between $\pm \sqrt{(T_e+T_i)/m_i}$ with the sheath angle coordinate $\varphi_w$ defined in Eq.~\eqref{eq:sheath_coordinate}. This is to conform to the sheath boundary conditions in Eq.~\eqref{eq:insulating_sheath}.

The velocity profile is initially symmetric in $\varphi$ while the toroidally symmetric density profile is perturbed by small fluctuations in order to trigger turbulence.

We define the source profile in Eq.~\eqref{eq:source} as
\begin{align}
    n_\mathrm{s}(R,Z) :=& n_{\text{prof}}(R,Z)D(R,Z), \label{eq:alignedX_damping}\\
    D(R,Z) :=& H_{\alpha}\left(\rho_{p,b}-\rho_p(R, Z)\right)
    \nonumber\\
   & H(Z-Z_X). \nonumber
\end{align}
We choose $\rho_{p,b} = 0.55$ for the source region, which is depicted as a dashed line Fig.~\ref{fig:flux}.
\subsection{The q-profile}
We follow the methods presented in~\cite{Wiesenberger2023FSA}
and define the geometric poloidal angle $\Theta$ as the field-line following
parameter around the O-point
\begin{align*}
\Theta = \begin{cases}
+\arccos\left[(R-R_O)/r\right] \text{ for } R \geq R_O \\
-\arccos\left[(R-R_O)/r\right] \text{ for } R < R_O 
\end{cases},
\end{align*}
with $r^2 := (R-R_O)^2 + (Z-Z_O)^2 $. We then have with $\vec B$ given by Eq.~\eqref{eq:magnetic-field}
 $B^\Theta = \vec B\cn\Theta = -(\psi_R (R-R_O) + \psi_Z (Z-Z_O))/(r^2R)$.
We can then directly integrate any field-line as
\begin{align*}
\frac{\d R}{\d\Theta} = \frac{B^R}{B^\Theta},\qquad
\frac{\d Z}{\d\Theta} = \frac{B^Z}{B^\Theta},\qquad
\frac{\d \varphi}{\d\Theta} = \frac{B^\varphi}{B^\Theta},
\end{align*}
from $\Theta=0$ to $\Theta=2\pi$. The safety factor results via
\begin{align}\label{eq:safety_factor}
q\equiv\frac{1}{2\pi}\oint \frac{B^\varphi}{B^\Theta} \d\Theta.
\end{align}
\begin{figure}[htbp]
    \centering
    \includegraphics[width = 0.45\textwidth]{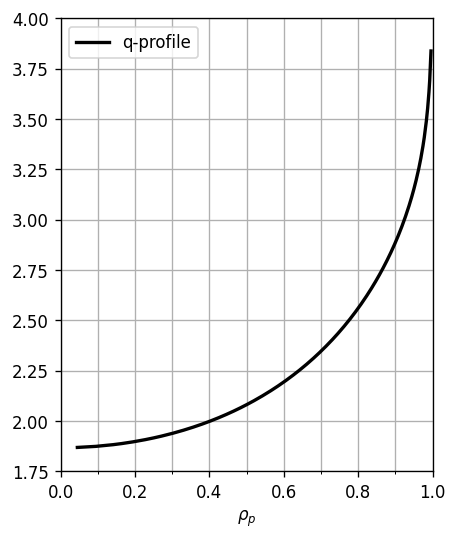}
    \caption{The q-profile as a function of the normalized poloidal flux $\rho_p$~\eqref{eq:rhop}. $q$ diverges at it approaches $\rho_p=1$ (the separatrix) but converges to a finite value $q\approx 1.9$ at $\rho_p=0$ (the O-point).
    }
    \label{fig:q-profile}
\end{figure}
Fig.~\ref{fig:q-profile} shows the q-profile of the chosen equilibrium. As expected the q-profile diverges at the separatrix situated at $\rho_p = 1$.
This is because $B^\Theta = 0$ at the X-point and thus the integration in Eq.~\eqref{eq:safety_factor} diverges. At the O-point around $\rho_p=0$ the q-profile converges
to a finite value $q\approx 1.9$. In the domain between $\rho_p = 0.4$ and $\rho_p = 0.9$ the value of $q$ lies between $2$ and $3$.


\subsection{A parameter scan}
We setup parameters for in total 12 simulations as two sets of 6 simulations each. The first
set uses $T_i = 0$ while the second set uses $T_i = T_e$. The 6 simulations within each set vary the dimensionless plasma resistivity $\eta$ Eq.~\eqref{eq:resistivity}, while keeping the plasma density $n_0 = 10^{19}$~m$^{-3}$ and
$\rho_s=1$ mm constant. This is achieved by changing the electron temperature $T_e$ (to set $\eta$) and the magnetic field strength $B_0$ (to keep $\rho_s\propto \sqrt{T_e}/B_0$ constant) as shown in Table~\ref{tab:parameters}. This results in a constant value for the plasma beta $\beta:= n_0 T_e / (B^2/(2 \mu_0)) = 10^{-4}$.
\begin{table}
\centering
\begin{tabular}{rrrrc}
\toprule
 & $B_0/$T & $T_e/$eV & $\omega_s^0 / $kHz & $\omega_s^1 / $kHz \\
$\eta$ &  &  &  &  \\
\midrule
1.00e-06 & 1.27 & 77.76 & 1.53 & 1.53 \\
3.00e-06 & 0.97 & 44.90 & 1.39 & 1.39 \\
1.00e-05 & 0.72 & 24.59 & 1.20 & 1.20 \\
3.00e-05 & 0.54 & 14.20 & 1.30 & 1.30 \\
1.00e-04 & 0.40 & 7.78 & 1.35 & 1.93 \\
3.00e-04 & 0.31 & 4.49 & 2.35 & 2.93 \\
\bottomrule
\end{tabular}

\caption{Parameters corresponding to varying the dimensionless plasma resistivity $\eta$ Eq.~\eqref{eq:resistivity} while keeping $n_0 = 10^{19}$m$^{-3}$ and $\rho_s=1$ mm constant. This results in constant $\beta = 10^{-4}$ and various $B_0$ and $T_e$ values.  The source strength parameter $\omega_s^0$ in Eq.~\eqref{eq:source} corresponds to $T_i = 0$ simulations while $\omega_s^1$ corresponds to $T_i=T_e$ simulations.  We select $B_0 \propto \eta^{-1/4}$ and $T_e \propto \eta^{-1/2}$.
}
 \label{tab:parameters}
\end{table}
The source strength parameter $\omega_s$ in Eq.~\eqref{eq:source} is constant for the duration of each simulation and chosen (differently for each simulation) such 
that the volume integrated source roughly matches the
total density flux out of the last closed flux-surface. 

We set the dimensionless parallel density diffusion necessary for numerical stability of the FCI scheme to a constant value $\nu_{\parallel,N} = 500$. The  perpendicular hyper-diffusion coefficients are set to $\nu_{\perp,N}=\nu_{\perp,U}=10^{-3}$.

The simulation domain is a rectangle in the $R$-$Z$ plane chosen such that the closed field line region as well as the SOL, wall and sheath regions are captured as displayed in Fig.~\ref{fig:flux}. It is important for the stability of the FCI scheme that the boundary of the wall region does not intersect the boundary of the simulation domain except at the sheath region.
The domain is symmetric in the $\varphi$ direction.   
The resolution is chosen as $192$ cells in $R$ and $336$ cells in $Z$ direction with $3$ polynomial coefficients in each cell in both $R$ and $Z$.
The number ratio $N_R/N_Z$ corresponds approximately to the aspect ratio of the simulation domain such that the grid cells are square in $R$-$Z$.
In $\varphi$ we choose $32$ planes. In total we thus have $576\cdot 1008 \cdot 32 \approx 2\cdot 10^7$ grid points.
Each simulation is run to roughly the same end time of $100\,000\ \Omega_{0}^{-1}$ with exact values displayed in Table~\ref{tab:endtimes}. The value $100\, 000$ is chosen as a compromise between a reasonable simulation wall time and a long enough, i.e. statistically significant, time series for our analysis in the following Section~\ref{sec:results}. 
\begin{table}
\centering
\begin{tabular}{rrrrr}
\toprule
 & \multicolumn{2}{r}{$T_i=0$} & \multicolumn{2}{r}{$T_i=T_e$} \\
 & $t_\mathrm{end}/\Omega_{i0}^{-1}$ & $t_\mathrm{end}/$ms & $t_\mathrm{end}/\Omega_{i0}^{-1}$ & $t_\mathrm{end}/$ms \\
$\eta$ &  &  &  &  \\
\midrule
1.00e-06 & 110400 & 1.81 & 111800 & 1.83 \\
3.00e-06 & 110200 & 2.38 & 111200 & 2.40 \\
1.00e-05 & 97500 & 2.84 & 88800 & 2.59 \\
3.00e-05 & 100000 & 3.83 & 100000 & 3.83 \\
1.00e-04 & 89165 & 4.62 & 100000 & 5.18 \\
3.00e-04 & 100000 & 6.82 & 99800 & 6.80 \\
\bottomrule
\end{tabular}

\caption{Simulation end times in units of $\Omega_{i0}^{-1}$ and in physical units reached after an equal amount of simulation time for all parameters. Simulations are run on $16$ NVidia V100 GPUs. 
}
 \label{tab:endtimes}
\end{table}
The end-time in units of ms is however different for each simulation and depends on the magnetic field strength corresponding to the chosen resistivity as depicted in Table~\ref{tab:parameters}. Since we keep $\rho_s\propto \sqrt{T_e}/B_0$ constant, changing the electron temperature $T_e$ yields a corresponding change in $B_0$ and thus $\Omega_{i0}$.

\subsection{Performance observations}
 The given resolution of $2\cdot 10^7$ grid points corresponds to an array size of $150$MB for each of the density, velocity and potential variables.
 With simulation data written to file at every $150\Omega_{i0}^{-1}$ the total file size of one simulation is about $500$GB.
The grid size is about a factor $5-100$ smaller than is currently used for (five-dimensional) gyro-kinetic simulations~\cite{Ku2018,Germaschewski2021,Mandell2022} but is of similar order of magnitude as other fluid-type simulation runs~\cite{Zholobenko2021,Giacomin2022}. 

Our simulations were run on $16$ NVidia V100 GPUs (equivalent to $4$ nodes on the M100 GPU cluster).
In Table~\ref{tab:timings} we present the average runtime in seconds per $\Omega_{i0}^{-1}$ for each simulation with the error being the standard deviation. These timings include the times for input/output and diagnostics but exclude the times for initialization and restarting of the code. 
Typically we achieve a computation time of $5-7$s per $\Omega_{i0}^{-1}$ but outliers at $4.6\pm 0.6$s and $8.3\pm 0.2$s exist. The differences may be due to slightly different viscosity parameters that we chose to stabilize some simulations and subsequent smaller or larger simulation time steps. The evaluation of a single right hand side of Eqs.~\eqref{eq:density} and \eqref{eq:parallel-mom} including solutions of all elliptic equations and evaluation of the parallel advection-diffusion terms takes about $0.20-0.25$ s in all simulations. The polarization equation~\eqref{eq:polarisation} is solved in typically $0.05$ s and less than $0.1$ s.
 The right hand side has to be evaluated 3 times per time step. 
\begin{table}
\centering
\begin{tabular}{rcc}
\toprule
 & $t_\text{comp}(T_i=0) / s$ & $t_\text{comp}(T_i=T_e) / s$ \\
$\eta$ &  &  \\
\midrule
1.00e-06 & 5.3 ± 0.7 & 5.3 ± 0.2 \\
3.00e-06 & 5.3 ± 0.8 & 5.3 ± 0.2 \\
1.00e-05 & 5.6 ± 1.0 & 6.6 ± 0.3 \\
3.00e-05 & 7.2 ± 0.2 & 7.4 ± 0.6 \\
1.00e-04 & 4.6 ± 0.6 & 5.4 ± 0.4 \\
3.00e-04 & 6.1 ± 0.4 & 8.3 ± 0.2 \\
\bottomrule
\end{tabular}

\caption{Average computational time per $\Omega_{i0}^{-1}$ in seconds. A runtime of $6$ s per $\Omega_{i0}^{-1}$ corresponds to a total simulation time of $7$ days for $100\,000\ \Omega_{i0}^{-1}$.
}
 \label{tab:timings}
\end{table}

As pointed out in our performance study~\cite{Wiesenberger2019Feltor} the observed code performance is bound by memory bandwidth and memory latencies. We emphasize that due to our structured grid approach our matrix-vector multiplications are almost as fast as vector additions since the matrix elements can be kept in cache. This and the major reduction in memory requirements that comes with it are the major benefits over unstructured grids. 
Of the total peak performance of $14\,400$ GB/s our implementation (of vector addition, matrix-vector multiplication, scalar products) reaches on average  $70$\%. 
We can compare this to the conventional Skylake partition on the Marconi cluster where one node has a theoretical peak bandwidth of $256$ GB/s of which our implementation on average (vector addition, matrix-vector multiplications, scalar products) achieves $183$ GB/s. With $16$ nodes we thus have an available throughput of $4096$ GB/s, which is a factor $3.5$ less than what is available on $4$ nodes on the M100 cluster. We see about a factor $3$ in practice, i.e. a runtime of $15$ s per $\Omega_{0}^{-1}$ for the $\eta=10^{-4}$ simulations and approximately $0.7$ s per right hand side evaluation.


\section{A study of resistivity and temperature} \label{sec:results}
In this Section we analyse the simulations previously setup in Section~\ref{sec:setup}. In Section~\ref{sec:visualisation} we show selected three-dimensional renderings of the magnetic field, plasma density and parallel current. Following this we establish the flux surface average in Section~\ref{sec:fsa} as a diagnostics tool for a numerical verification of the simulations in Section~\ref{sec:verification}. We focus on the parallel acceleration in Section~\ref{sec:acceleration} and mass and energy confinement in Section~\ref{sec:confinement}.
\begin{figure*}[htbp]
    \centering
    \includegraphics[width = \textwidth]{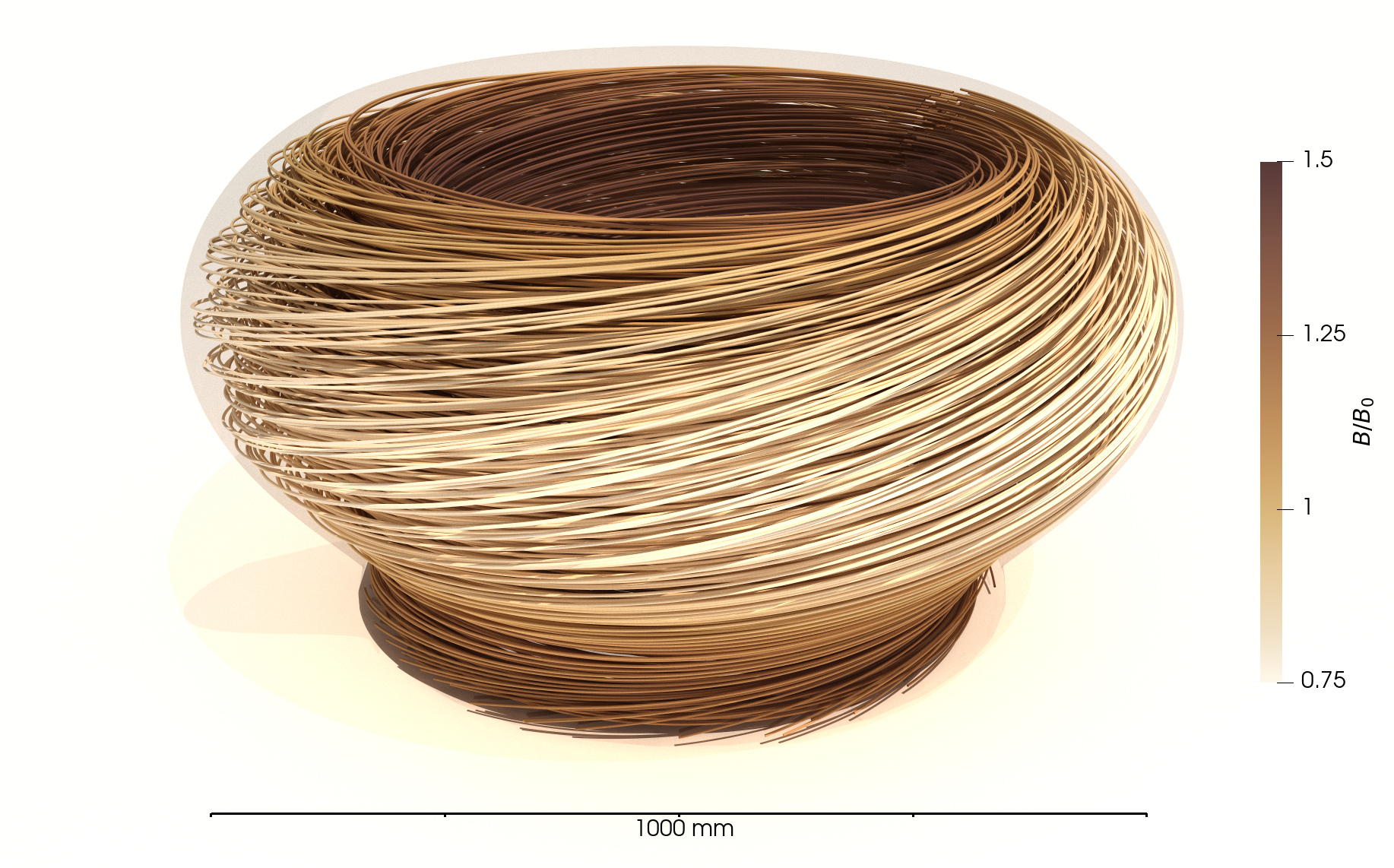}
    \caption{Streamlines of the magnetic field vector $\vec B$ integrated and visualised in ParaView~\cite{paraview}. One low-opacity iso-contour of $\rho_p=1.10$ is plotted (corresponding to $\psi_p = 4$). The positive $\vec B$ direction is clockwise if viewed from above and the field-line winding is left-handed. $\vec B\times\vn\vec B$ points towards the magnetic X-point and we have a favourable drift direction.
    }
    \label{fig:magnetic_field}
\end{figure*}
\begin{figure*}[htbp]
    \centering
    \includegraphics[width = \textwidth,trim=79 0 77 0px,clip]{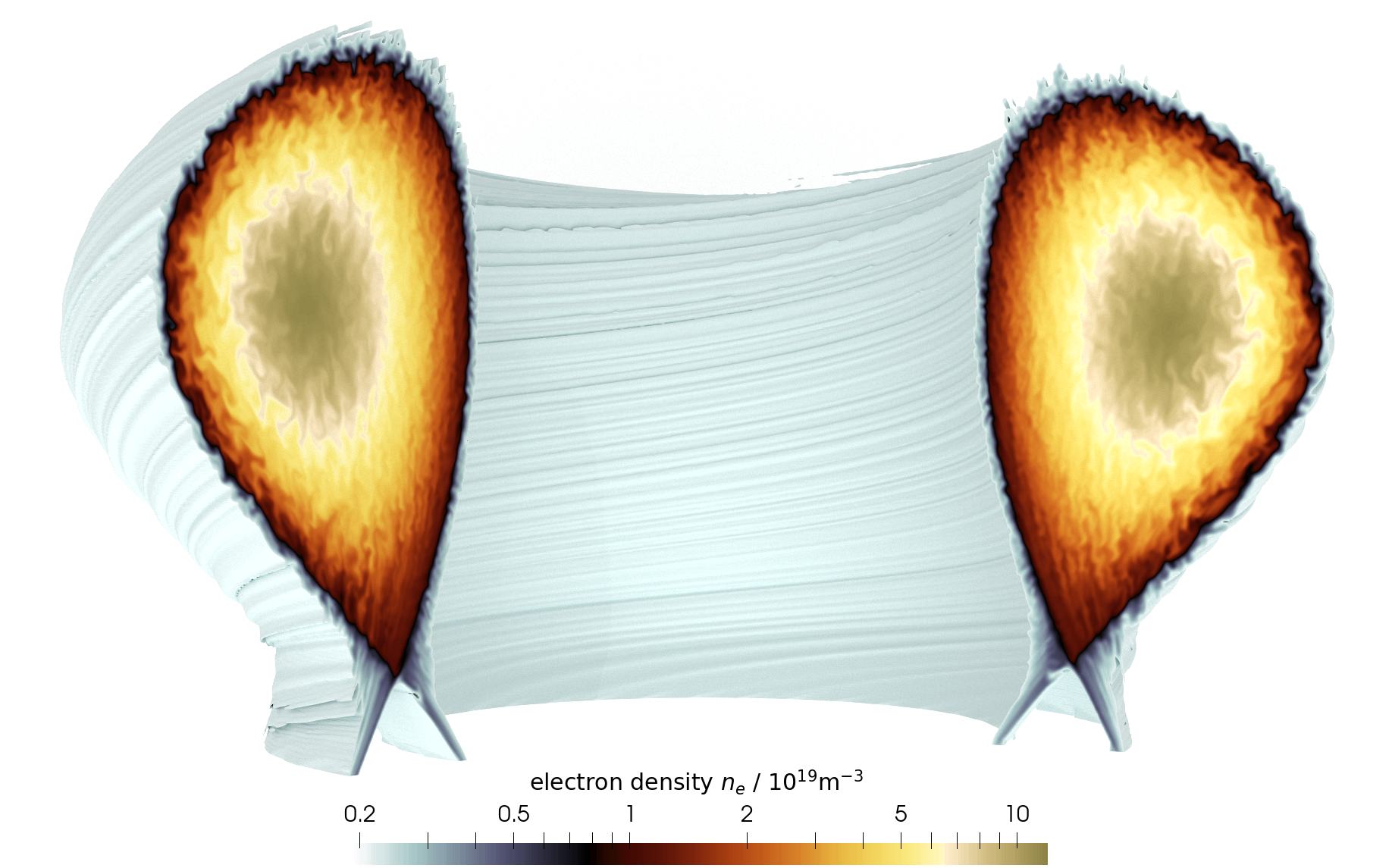}
    \caption{The electron density $n_e$ at $5.18$ms for $\eta=10^{-4}$, $T_i = T_e = 7.8$eV and $B_0 = 0.4$T. We show an iso-volume of $n_e/n_0 \geq 0.22$ and choose a wave colourmap constructed with the ColorMoves tool from \cite{sciviscolor} mapped to logarithmic density values. The three colour regions (blue-grey, red-yellow and brown-grey) roughly coincide with the three regions scrape-off layer, edge and core/source region (cf. Fig.~\ref{fig:fluxb})
    }
    \label{fig:electron_density}
\end{figure*}

\begin{figure*}[!htbp]
    \centering
    \includegraphics[width = \textwidth,trim=79 0 77 0px,clip]{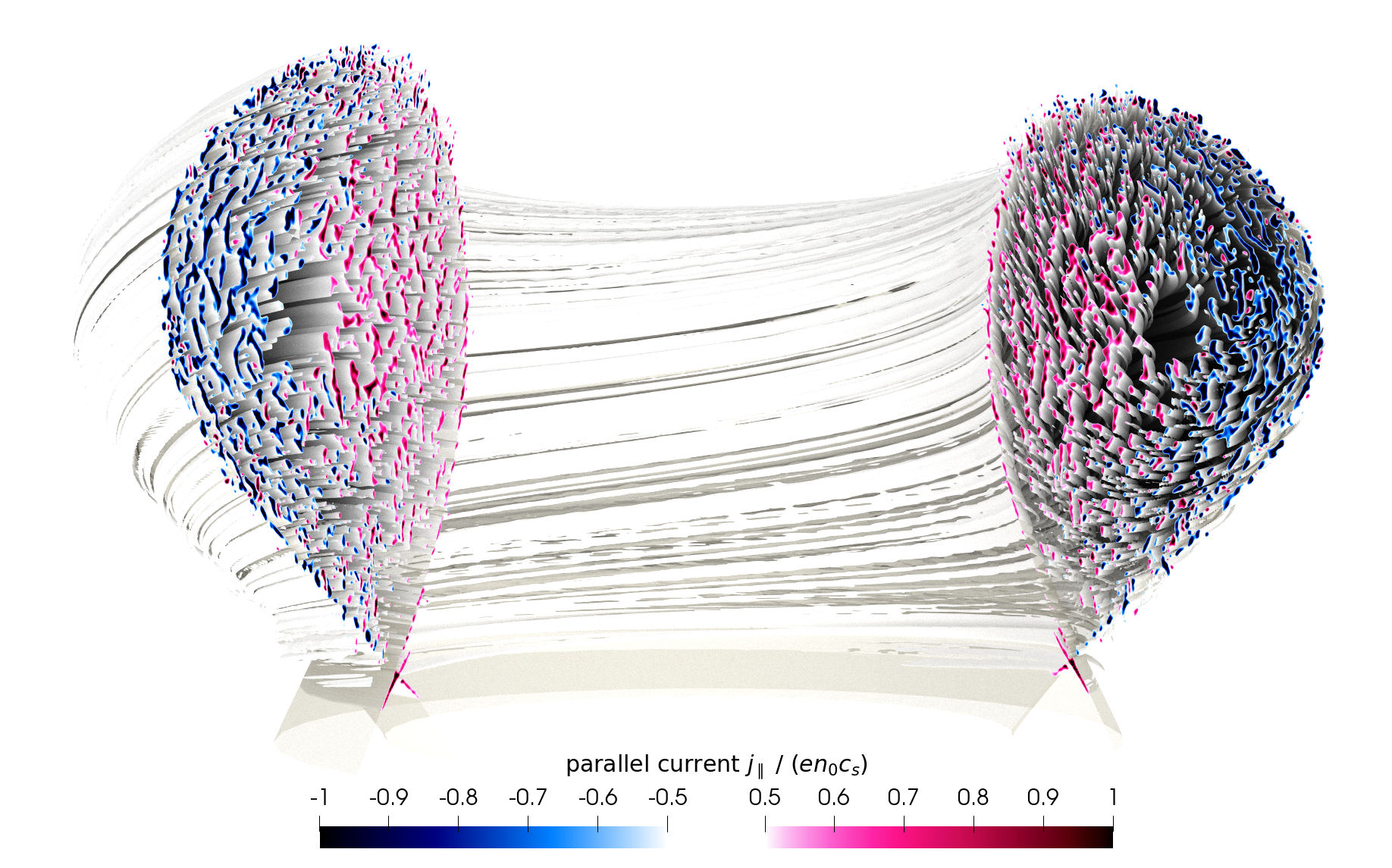}
    \caption{The parallel electric current $j_\parallel/(en_0c_s)$ at $5.18$ms for $\eta=10^{-4}$, $T_i = T_e = 7.8$eV and $B_0 = 0.4$T. 
    We plot two isovolumes $j_\parallel \leq -0.5en_0c_s$ and $j_\parallel \geq 0.5en_0c_s$. The colour-scale is cut at $-1$ and $1$ respectively. A translucent contour of the separatrix $\psi_p=0$ is shown. Current mainly flows in field-aligned tubes. Each tube has a typical extension of $5$ mm and thus carries approximately $25 en_0c_s\rho_s^2 \approx 2.5$ A. 
    }
    \label{fig:current}
\end{figure*}

\subsection{Three-dimensional visualisations} \label{sec:visualisation}
Here, we present three-dimensional renderings of the magnetic field and the density and parallel current of the $\eta=10^{-4}$, $T_i=T_e$ simulation. The ParaView visualisation toolkit~\cite{paraview} is used and all results are rendered on a NVidia RTX3090 card.
In order to render the $\varphi$ direction smoothly we face a challenge due to the low resolution of only $32$ toroidal planes. To solve the issue we temporarily extend the available data to $384$ toroidal planes by interpolating along the magnetic field lines with the methods presented in~\cite{Wiesenberger2023FSA}.
This allows for a smooth visualisation of the field-line following structures.
 \subsubsection{Magnetic field}
We begin by showing a three-dimensional rendering of the magnetic streamlines in Fig.~\ref{fig:magnetic_field}.
We use the streamline tracer module in ParaView~\cite{paraview} to integrate magnetic field lines of Eq.~\eqref{eq:magnetic-field}
and visualise with the OptiX path tracer using $1000$ progressive passes of the ray tracer.  A low-opacity iso-contour of $\rho_p = 1.10$ is plotted in order to remove spurious visualisation artifacts. A light source is placed approximately in the upper right corner of the viewing space and a flat, white, non-opaque surface is placed at $Z=-450$ mm in order to aid the lighting of the scene that has an otherwise dark grey background behind the camera. The colour scale is chosen from~\cite{Samsel2019,sciviscolor} and is used to represent the magnetic field strength following the "dark-is-more" bias~\cite{Schloss2018} for easier interpretation. 

Using ray tracing gives the impression of a photo-realistic image of the magnetic streamlines with an enhanced depth-perception and an easy distinction of inner vs outer streamlines. At the same time, shadows in general and in particular the shadows falling on the "floor" in the lower left corner of the image are visual enhancements that have no actual physical reality.

The streamlines follow a left handed winding with the positive $\vec B$ direction clockwise if viewed from the top. Only magnetic streamlines in the scrape-off layer are visible, which originate at the numerical divertor at the bottom. The magnetic field strength is clearly higher on the interior side (high field side) than on the outside (low field side) following the general $1/R$ dependence of Eq.~\eqref{eq:magnetic-field-strength}.  As mentioned in Section~\ref{sec:magnetic} the $\vec B\times\vn\vec B$ direction points towards the magnetic X-point and we have a favourable drift direction. 
\subsubsection{Electron density}
The electron density is depicted in Fig.~\ref{fig:electron_density}.
Here, we create an iso-volume for $n_e/n_0 \geq 0.22$ between the angles $0$ and $250^\circ$ in the $\varphi$ direction. This enables the viewer to see both the field-alignment in the scrape-off layer as well as a cross-section of the perpendicular turbulence in the edge and core regions.

As a colour-scale we create a three-colour map with the help of the ColorMoves tool~\cite{sciviscolor} with transitions at $0.8$ and between $6$ and $7$.
The three colours can be interpreted as visualisations of scrape-off layer (grey-blue), edge (red-yellow) and core (brown-grey). Here, the core is the region where our particle source is active (cf. the dashed line in Fig.~\ref{fig:flux}).
 The motivation for choosing such a colour scale for the density is the large data volume spanning almost two orders of magnitude with relatively small fluctuations on top.
 We follow the colour-name variation metric as promoted by~\cite{Szafir2021} as opposed to a colour scale that varies purely in luminance say. The latter would help to visually establish order, that is darker regions correspond to higher density values. However, we found no single-colour scale that could span the large data volume while maintaining good colour discriminability. We thus sacrifice some uniformity and order at the transition points in favour of a higher discriminative power, i.e. a higher amount of distinct colours. As a result
 it is not directly intuitive which colour corresponds to higher density values without consulting the colourmap, however, the turbulent structures in the core and edge as well as the filamentary structures in the scrape-off layer are highly visible.
 
As was done in Fig.~\ref{fig:magnetic_field} we use the OptiX path tracer in ParaView with $1000$ passes to render the final image. As a lighting source we choose a large radius source directed from below the camera to eliminate sharp shadows and increase the contrast between the field-aligned structures in the scrape-off layer. We place a white plane behind the iso-volume (which the camera looks onto) and a light grey coloured background behind the camera. This
achieves a uniformly lit scene.

The scene itself shows the largest turbulent fluctuations in the core and edge regions on the low field side, especially at the outboard midplane. Fluctuations on the high field side are smaller in perpendicular extension. This points towards a ballooning mode. Further, we notice that fluctuations are highly elongated vertically at the top of the domain as well as at the bottom around the X-point both in the edge as well as the scrape-off layer. The scrape-off layer fluctuations appear field aligned judging from the form of the contours in between the two poloidal planes.
\subsubsection{Parallel current}

The next visualisation is the parallel current $j_\parallel = e(N_i U_{\parallel,i} - en_e u_{\parallel,e})$ in Fig.~\ref{fig:current}.  We create two separate iso-volumes for $j_\parallel$: one for $j_\parallel / (en_0c_s )\geq 0.5$ and one for $j_\parallel / (en_0c_s) \leq 0.5$. Here, we use $c_s = \sqrt{T_e/m_i} = 4.64\cdot10^4$ m/s. Two separate colourmaps are chosen for each region; a blue one for the negative and a red one for the positive values.
Both colourmaps begin at $\pm 0.5$ and are truncated at $\pm 1$ (actual values lie between $\pm 4.7$).

We choose a similar setup as for the density rendering, i.e. a white plane behind the scene with a light grey background behind the camera. A large radius headlight is placed at the camera to illuminate the scene. Again, ray tracing is used to render the final image. In order to guide the viewer we plot a low-opacity iso-contour of $\rho_p = 1$ (the separatrix). 

The resulting image highlights the localized "field-aligned tubes" in which current flows in the simulation. These tubes have a typical extension of about $5$ mm and thus carry a current of approximately $25 en_0c_s\rho_s^2 \approx 2.5$ A. It is further visible that the current is positive (flow direction clockwise viewed from above) mainly on the high-field side and negative mainly on the low-field side. However, a couple of individual current lines of the opposite signs are discernible in either region. Few current lines exist in the scrape-off layer and only close to the separatrix.
\subsection{The flux surface average - profiles and fluctuations} \label{sec:fsa}
\begin{figure}[htbp]
    \centering
    \includegraphics[width = 0.4\textwidth]{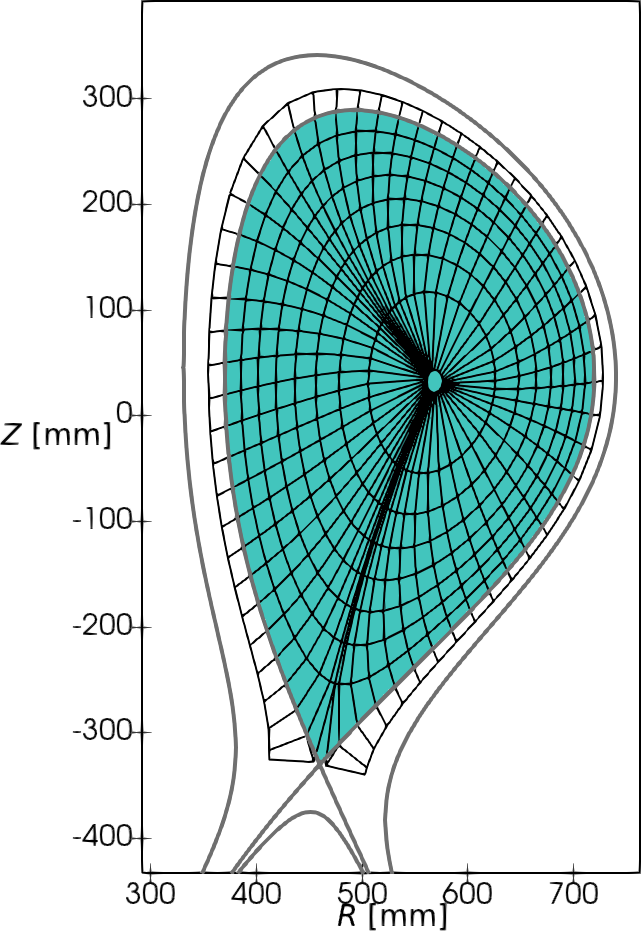}
    \caption{The flux-aligned grid (with $20\times$ reduced resolution to see the grid points) used for the computation of flux-surface averages and flux-volume integration. The closed field line region $\Omega$ for the verification is shown in blue and contains a volume of $0.5$ m$^3$. The grid allows for a definition of a flux-surface average outside the separatrix.
    }
    \label{fig:integration-region}
\end{figure}

Before we can turn to a verification exercise of our simulations we first need to establish appropriate integration routines. More specifically we here want to compute so called flux-surface averages and integrals. The flux-surface average is defined as a differential volume average according to~\cite{haeseleer}:
\begin{align} 
\RA{ f }(\psi_p) :=& \frac{\partial}{\partial v} \int  f\dV, \label{eq:fsa} \\
v(\psi_{p,0}) :=& \int H(\psi_p(R,Z) - \psi_{p,0}) H(Z-Z_X) \dV, \label{eq:vol}
\end{align}
where $H(x)$ is the  Heaviside function.
In order to accurately integrate Eqs.~\eqref{eq:fsa} and \eqref{eq:vol} we use the methods described in~\cite{Wiesenberger2023FSA}. The
first step is to construct a flux aligned coordinate system as we show in Fig.~\ref{fig:integration-region}.

There are several numerical pitfalls that should be considered when numerically constructing a flux-aligned grid. 
As pointed out in Reference~\cite{Wiesenberger2018,Wiesenberger2023FSA} the volume element in flux-aligned grids diverges and care must be taken when constructing such grids close to the X-point. This is especially true if the flux-surface average of the separatrix (or a surface close to it) is to be computed. We follow~\cite{Wiesenberger2017,Wiesenberger2018} for the construction of our grid.

In flux-aligned coordinates $\eta$, $\zeta$, $\varphi$ the flux-surface average simplifies to
\begin{align}\label{eq:fsa_aligned}
 \RA{f} = \frac{1}{2\pi\oint \sqrt{g}\d\eta } \oiint_0^{2\pi} f(\zeta(\psi_{p}),\eta,\varphi) \sqrt{g}\d\eta\d \varphi,
\end{align}
where $\sqrt{g}$ is the volume element in the flux aligned coordinate system.

The numerical integration in the $\varphi$ direction is straightforward. 
The resulting toroidal average can be interpolated onto the flux-aligned grid displayed in Fig.~\ref{fig:integration-region}. Then, Eq.~\eqref{eq:fsa_aligned} can be used to compute the flux surface average. Since the grid in Fig.~\ref{fig:integration-region} exists also outside the last closed flux surface, we can use Eq.~\eqref{eq:fsa_aligned} to compute flux-surface averages in the scrape-off layer as well.

\begin{figure*}[htbp]
    \centering  
    \includegraphics[width = \textwidth]{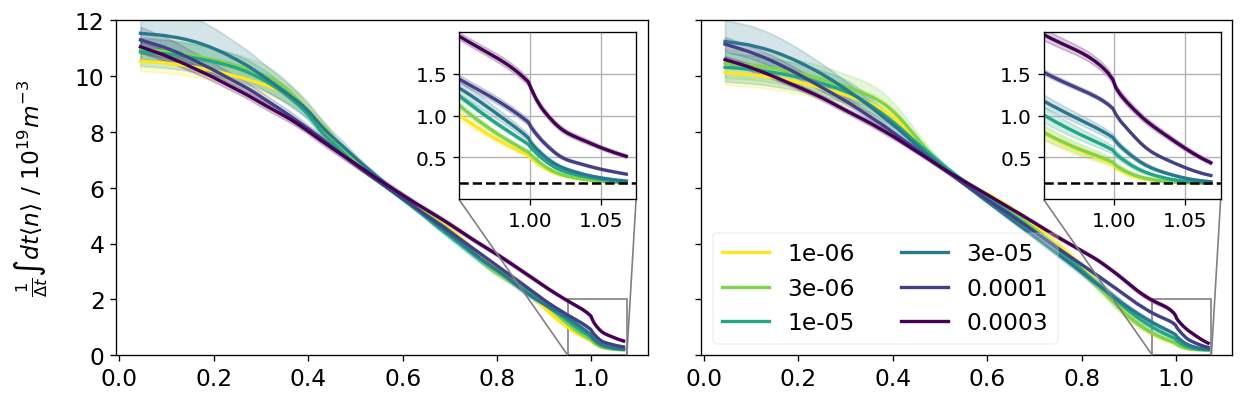}
    \includegraphics[width = \textwidth]{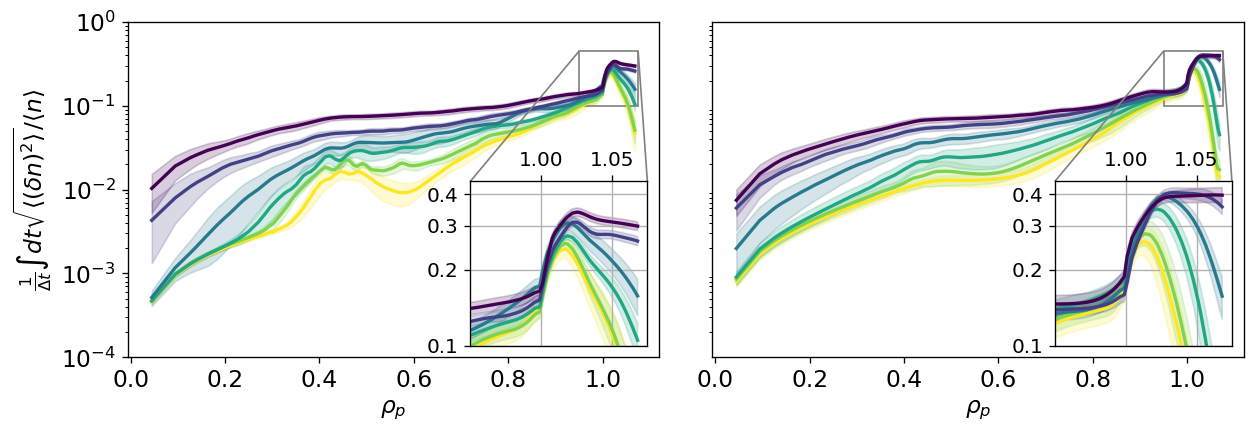}
    \caption{The time averaged density profiles (top) and the relative fluctuation amplitudes (bottom) for $T_i=0$ (left) and $T_i=T_e$ (right) as a function of $\rho_p$ Eq.~\eqref{eq:rhop}. The separatrix corresponds to $\rho_p = 1$. The edge and scrape-off layer regions $0.95<\rho_p<1.075$ are shown enlarged.
    }
    \label{fig:ne-profile}
\end{figure*}
In Fig.~\ref{fig:ne-profile} in the top half we show the flux-surface averages of the density $\RA{n_e}$ as a function of $\rho_p$ defined in Eq.~\eqref{eq:rhop}. In fact, we show a time averaged $\RA{n_e}$ profile for all simulations. The average profiles for $T_i=0$ and $T_i=T_e$ are visibly very similar. For the high resistivity simulations $\eta=3\cdot10^{-4}$ and $\eta=10^{-4}$ (both $T_i=0$ and $T_i=T_e$) the average profile appears linear in $\rho_p$ up to the separatrix at $\rho_p = 1$. The remaining simulations
have accumulated density in the core at about $\rho_p < 0.4$. This is the region where the source $S_{n_e}$ is active and continuously increases the density, which also translates to a large variation amplitude in the core. The edge and scrape-off layer at $0.95<\rho_p<1.075$ are shown enlarged. The density on the separatrix increases with resistivity from $0.5\cdot10^{19}$~m$^{-3}$ to about $1.5\cdot10^{19}$~m$^{-3}$ for both $T_i=0$ and $T_i=T_e$ simulations. Afterwards, in the scrape-off layer at $\rho_p > 1$ the density sharply drops. Notice that the black dashed line in the enlarged region signifies the minimum density $n_{e,\min} = 0.2\cdot10^{19}$~m$^{-3}$ in Eq.~\eqref{eq:minimumne}.
The average densities thus cannot reach below the artificially enforced lower boundary.
It may be preferable to run simulations
with lower $n_{e,\min}$ to study if the lower resistivity simulations converge at a different value, however then also the parallel viscosities $\nu_{\parallel}$ must be adapted in Eq.~\eqref{eq:restriction} in order to not also lower the CFL condition.

We define the relative fluctuation amplitude as
\begin{align}
    \sigma_{n_e}(\rho_p, t) := \frac{\sqrt{\RA{ (n_e - \RA{n_e})^2}}}{\RA{n_e}}.
\end{align}
In the lower part of Fig.~\ref{fig:ne-profile} we show the time averaged $\sigma_{n_e}$ for our simulations. 
Again, both the $T_i=0$ and $T_i=T_e$ simulations exhibit similar behavior. The fluctuation levels in the core region lie between $10^{-3}$ and $10^{-2}$ at the smallest $\rho_p$ where higher resistivity corresponds to higher fluctuation levels. The relative fluctuation amplitudes increase for all simulations to about $15\%$ at the separatrix. There is sharp increase in fluctuations for $\rho_p>1$ to a maximum of $35\%$ for $T_i=0$ and $40\%$ for $T_i=T_e$, visible in the enlarged regions of Fig.~\ref{fig:ne-profile}. Furthermore, between about $1<\rho_p<1.01$ the amplitudes for all simulations overlap before they decrease again at about $\rho_p = 1.02$. The small resistivity simulations decrease furthest in fluctuation amplitudes.

The observed radial profiles for density and its fluctuations can be tentatively compared with~\cite{Tatali2021} where a non-isothermal drift-fluid model is used to simulate the turbulent dynamics in a limiter configuration using buffer regions to exclude the core region from the simulation domain. There, the fluctuation level at the separatrix peaks only for small resistivities. Furthermore the separatrix densities are highest for smallest resistivities instead of largest resistivities as in our case. This is likely a consequence of how the source term $S_N$ depends on $\eta$. In the present case the source strength is adapted (see Table~\ref{tab:parameters}) such that the density profiles across simulations remain similar, while~\cite{Tatali2021} keeps an absolute source strength. 

\subsection{Verification of conservation laws}\label{sec:verification}
With a reliable way to compute flux-surface averages and volume integration we can now turn to defining a suitable error norm for a numerical verification. First, we again emphasize that due to the turbulence nature of our simulations, we cannot show pointwise convergence. In fact, in Reference~\cite{Wiesenberger2019Feltor} it
is shown that even computational errors on the order of machine precision in two-dimensional simulation exponentially increase to order one within a short period of time. This means that the occasionally used method of manufactured solution~\cite{Dudson2016,Tamain2016,Halpern2016,Giacomin2022} is not suitable for verifying simulation behaviour on a long timescale. We here therefore 
follow a different strategy where we compute the volume and time integrated error of conservation laws.

Assume  that our model equations in Section~\ref{sec:model} allow for a local analytical balance equation of the form
\begin{align}
    \sum_i t_i (R,Z,\varphi,t) = 0
\end{align}
that is a sum of individual terms $t_i$ balances to zero.
First, we define a time average via
\begin{align} \label{eq:time-average}
    \langle t_i\rangle_t :=  \frac{1}{\Delta t} \int_{t_0}^{t_1} t_i (R,Z,\varphi,t) \d t
\end{align}
The time interval $[t_0, t_1]$ in Eq.~\eqref{eq:time-average} will in the following Section~\ref{sec:mass-conservation} be manually defined for each simulation by identifying a saturated turbulence state.

Under a further volume integration we can convert the $t_i$ to
\begin{align} \label{eq:definition_ti}
  T_i :=& \left\langle \int_\Omega t_i(R,Z,\varphi,t) \d V \right\rangle_t
\end{align}
The spatial integration region  in Eq.~\eqref{eq:definition_ti} is chosen as the closed field line region $ \Omega :=\{ (R,Z,\varphi) : Z > Z_X \wedge \rho_p(R,Z) < 1\} $ and shown in colour in Fig.~\ref{fig:integration-region}. Note that once we have the flux-surface average $\RA{t_i}$ on a sufficiently fine grid in $\psi_p$ we can integrate
\begin{align*} 
    \int_\Omega t_i dV  = \int \RA{t_i} \d v = \left(\frac{\d v}{\d\psi_p}\right)^{-1}\int \RA{t_i}(\psi_p)  \d v
\end{align*}

We then have $\sum_i T_i = 0$ analytically, however, numerically due to discretization errors we usually have
\begin{align} \label{eq:global-error}
\sum_i T_i^{\text{num}} = E
\end{align}
where $E$ is the total numerical error and $T_i^\text{num}$ is the numerical result given
by the discrete version of Eq.~\eqref{eq:definition_ti} computed by storing the individual $t_i^\text{num}$ in memory during a simulation. We would consider the conservation law well fulfilled numerically, if $E$ is small compared to the $T_i^\text{num}$.

The error $E$ consists of the contributions $E_i$ of the errors of each
individual term $E_i =  T_i^\text{num} - T_i$, i.e. $E = \sum_i E_i$. We are interested in the
error for each term, however, given $E$ we a priori cannot deduce $E_i$. In order to get an error estimate nevertheless, we here assume that the error contribution $E_i$ of
each term is determined by its magnitude $|T_i^\text{num}|$.
We introduce the relative global error
\begin{align} \label{eq:rel-global-error}
    \varepsilon := \frac{E}{\sum_i |T_i^\text{num}|}
\end{align}
with which we can define
\begin{align} \label{eq:error-ti}
    E_i := \varepsilon |T_i^\text{num}|
\end{align}
The corrected terms should read
\begin{align} \label{eq:corrected-ti}
    T_i^\text{corr} := T_i^\text{num} - E_i
\end{align}
It is easy to see that
\begin{align}
    \sum_i T_i^\text{corr} = 0
\end{align}
An interpretation of $\varepsilon$ is to signify "the importance of a physical effect on the global dynamics that was not captured by the numerics". In this sense any error below $1\%$ can be considered excellent, while anything above merits further discussion.

\begin{figure*}[htbp]
    \centering
    \includegraphics[width = \textwidth]{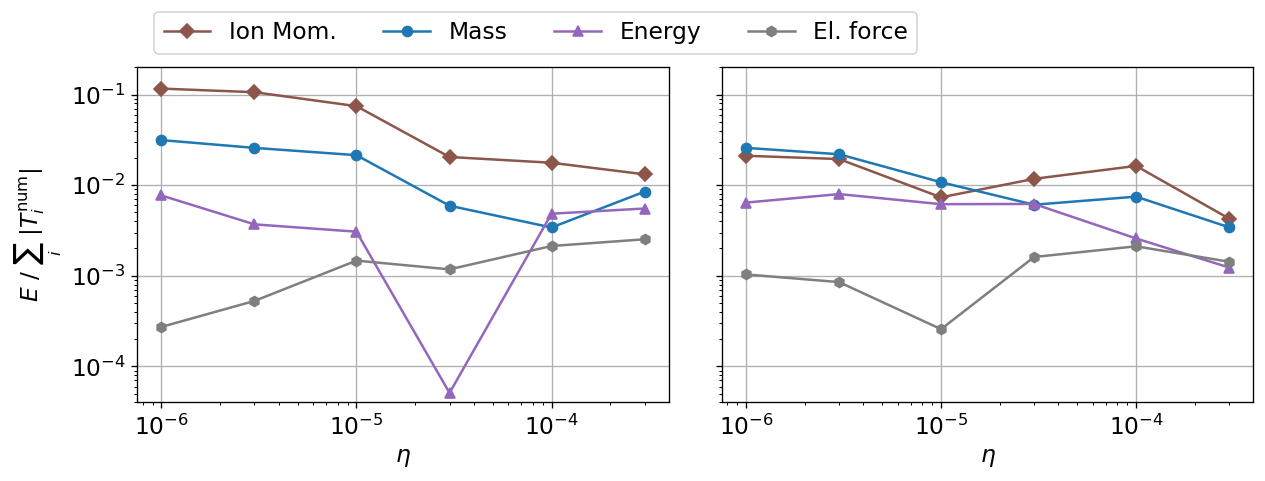}
    \caption{The relative global errors as defined by Eq.~\eqref{eq:rel-global-error} of the terms in the mass conservation in Section~\ref{sec:mass-conservation}, the energy theorem in Section~\ref{sec:energy}, the parallel momentum balance in Section~\ref{sec:parallel-mom} and the electron force balance in Section~\ref{sec:parallel-force} for $T_i=0$ (left) and $T_i=T_e$ (right).
    }
    \label{fig:relative-errors}
\end{figure*}

 We now analyse the mass conservation in Section~\ref{sec:mass-conservation}, the energy theorem in Section~\ref{sec:energy}, the parallel momentum balance in Section~\ref{sec:parallel-mom} and the electron force balance in Section~\ref{sec:parallel-force}. The resulting relative global errors are presented in Fig.~\ref{fig:relative-errors}.

\subsubsection{Mass conservation} \label{sec:mass-conservation}
The electron density equation~\eqref{eq:density} directly yields the particle conservation
\begin{align} \label{eq:mass_theorem}
  \frac{\partial}{\partial t} n_e
  + \nc\vec{ j_{n_e}}
  -  \Lambda_{n_e} - S_{n_e} = 0
\end{align}
with
\begin{align}
  \vec j_{n_e} =& \vec j_{n_e,E} + \vec j_{n_e,C} + \vec j_{n_e,\parallel} + \vec j_{n_e,A} \label{eq:mass-flux}\\
  \Lambda_{n_e} =& \Lambda_{n_e,\perp} + \Lambda_{n_e,\parallel}
\end{align}
where we split the density flux into the \ExB{} flux $\vec j_{n_e,E} := n_e \bhat \times \nabla \phi / B$, the curvature flux $\vec j_{n_e,C} = -n_e T \vec K/e - m_en_eu_{\parallel,e}^2 \vec \KK$, parallel flux $\vec j_{n_e,\parallel} = n_e u_{\parallel,e} \bhat$ and magnetic flutter flux $\vec j_{n_e,A}  = n_e u_{e,\parallel} \bperp$. The diffusive part consists of $\Lambda_{n_e,\perp} = -\mu_{n_e,\perp} \Delta_\perp^2n_e$ and $\Lambda_{n_e,\parallel} = \mu_{n_e,\parallel} \Delta_\parallel n_e$.


\begin{figure}[htbp]
    \centering
    \includegraphics[width = 0.49\textwidth]{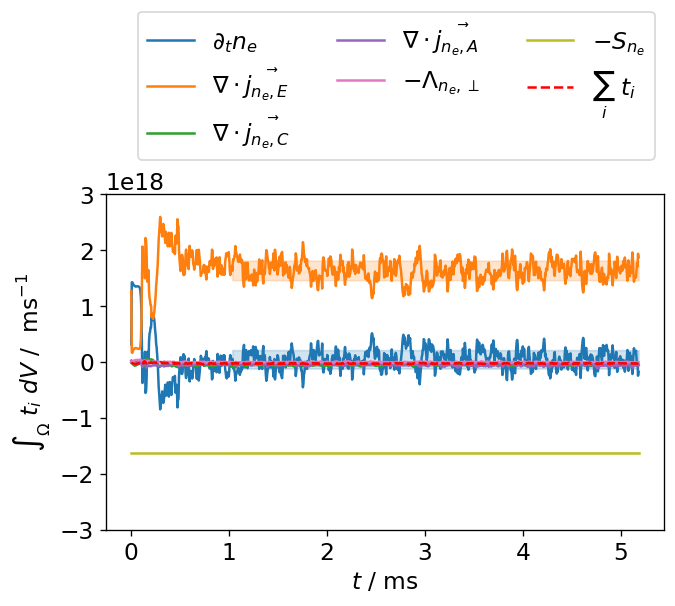}
    \caption{The time evolution of volume integrated terms in the mass conservation equation for $T_i=T_e$ and $\eta=10^{-4}$. 
    The length of the shaded regions signifies the time interval which we consider for our statistics while the widths signify the standard deviations within that region. }
    \label{fig:mass-conservation-lcfs}
\end{figure}
In Figure~\ref{fig:mass-conservation-lcfs} we plot the volume integrated terms of the mass conservation \eqref{eq:mass_theorem} as a function of time for the $T_i=T_e$ and $\eta=10^{-4}$ simulation. 
First, notice that $\RA{ \nc \vec j} = \frac{\d }{\d v} \RA{\vec j \cn v}$~\cite{haeseleer} and thus
\begin{align}\label{eq:div-equals-flux}
    \int_\Omega \nc \vec j \d V  = \int_{\partial\Omega} \vec j \cdot \vec{\dA} = \RA{\vec j \cn v}|_{\rho_p=1},
\end{align}
i.e. the volume integral of divergences equals the total flux out of the last closed flux surface  or the average radial flux.
We immediately see that the two largest actors in this figure are the \ExB{} flux $\RA{ \vec j_E\cn v}$ on the last closed flux surface and the density source $\int S_{n_e} \dV $, which is constant throughout the simulation.
The time derivative of the total mass fluctuates around zero. Note that the remaining terms including the error given by the sum of all terms $\sum_i t_i$ are too small to be visibly different from zero in the plot.

Further, notice that the flux surface average $\RA{ \nc (j_0 \bhat)} = \frac{\d }{\d v} \RA{j_0 \bhat \cn v} = 0$ vanishes for any parallel current $j_0\bhat$.
Any deviation from zero is thus purely numerical. This applies in particular to the terms $\nc \vec j_\parallel$ and $\Lambda_{n_e,\parallel}$ in Eq.~\eqref{eq:mass_theorem}. In our recent work in~\cite{Wiesenberger2023FCI} we individually study the deviations from zero in those terms and find them
to be negligibly small. We will thus here and in the following ignore parallel terms accepting that they may contribute to the errors visible in Fig.~\ref{fig:relative-errors}.

\begin{figure*}[htbp]
    \centering
    \includegraphics[width=\textwidth]{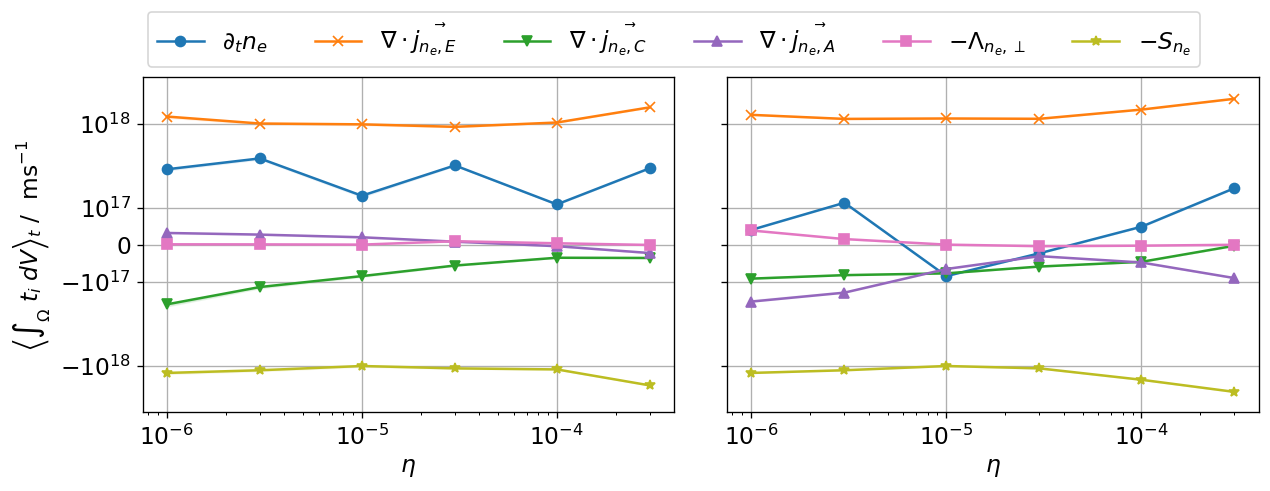}
    \caption{The mass conservation equation \eqref{eq:mass_theorem}: volume integrated and time averaged terms Eq.~\eqref{eq:definition_ti} with error bar Eq.~\eqref{eq:error-ti}  for $T_i=0$ (left) and $T_i=T_e$ (right). The error bars are too small to be visible in the plot and are separately shown in Fig.~\ref{fig:relative-errors}.
    }
    \label{fig:mass-importance}
\end{figure*}

From the \ExB{} flux in Fig.~\ref{fig:mass-conservation-lcfs} we manually identify a time interval where fluctuations appear around a constant average. We do this for all $12$ simulations. 
This allows us to identify suitable $t_0$ and $t_1 = t_{\mathrm{end}}$
in Eq.~\eqref{eq:definition_ti} and thus we can compute the relative global error in Eq.~\eqref{eq:rel-global-error}.  We plot the corrected terms~\eqref{eq:corrected-ti} together with error bar from Eq.~\eqref{eq:error-ti} in Fig.~\ref{fig:mass-importance}.
The left plot shows simulations with $T_i=0$ for the various resistivities $\eta$ and the right plot shows corresponding simulations with $T_i=T_e$. We can immediately confirm that the \ExB{} flux as well as the source term are the largest terms for all simulations while the time derivative follows with lesser importance. Note here that the density source strength $\omega_s$ in $S_{n_e}$ in Eq.~\eqref{eq:source} was chosen differently for each simulation.
The magnetic flutter term as well as the curvature flux and the perpendicular  diffusion terms have negligible importance on the evolution of the global mass balance. We emphasize that this does not necessarily imply negligible importance on the local dynamics just that the volume integrated mass balance is unaffected.

The relative errors in the terms are invisible in this plot, which is why we separately plot these in Fig.~\ref{fig:relative-errors}. There we see that the relative error of the terms in the mass conservation is at an excellent maximal $3$\% for all simulations and below $1$\% for simulations with $\eta > 10^{-5}$. 

\subsubsection{Energy theorem} \label{sec:energy}
The terms of the energy theorem are
\begin{align} \label{eq:energy_theorem}
\partial_t \mathcal E +
\nc \vec j_{\mathcal E}
- \Lambda_{\mathcal E}
-  S_{\mathcal E}
-  R_{\mathcal E} = 0
\end{align}
with 
\begin{align} \label{eq:energy_conservation}
  \mathcal{E}= & T_e n_e \ln{(n_e/n_{e0})} + T_i N_i\ln{(N_i/n_{e0})}
  \nonumber\\
 & +\frac{1}{2}\mu_0\left(\nperp A_\parallel\right)^2
   +  \frac{1}{2} m_i N_i u_E^2  \nonumber\\
   & +\frac{1}{2} m_e  n_e u_{\parallel,e}^2
  +\frac{1}{2} m_i  N_i U_{\parallel,i}^2,
  \end{align}
  \begin{align}
  \vec j_{\mathcal E} =& \sum_s \left[
  \left(T\ln (N/n_{e0}) + \frac{1}{2}m U_\parallel^2 + q\psi \right)\vec j_N \right]
  \nonumber\\
  &+ \sum_z \left[\frac{m}{q} NU_\parallel^2\KK + T NU_\parallel \left(\bhat + {\vec b}_\perp\right)\right],
    \end{align}
  \begin{align}
  \Lambda_{\mathcal E} =&  \sum_s \left[\left( T\left( 1+\ln{(N/n_{e0})}\right) + q\psi + \frac{1}{2} mU_\parallel^2 \right)
  \Lambda_N \right]  \nonumber\\
  &+m NU_\parallel\Lambda_U 
  \end{align}
  \begin{align} \label{eq:energy-source}
  S_{\mathcal E} =&  \sum_s  \left[ \left(T\left( 1+\ln{(N/n_{e0})}\right) +q\psi - \frac{1}{2} m U_\parallel^2 \right)S_{N}\right]
  \end{align}
  \begin{align}
R_{\mathcal E} =&  -\eta_\parallel e^2 n_e(U_{\parallel,i}-u_{\parallel,e})(N_iU_{\parallel,i} - n_eu_{\parallel,e}).
\end{align}
where in the energy flux $\vec j_{\mathcal E}$
we neglect terms  containing time derivatives
of the electric and magnetic potentials and we sum over all species.
The energy density $\mathcal E$ consists of the Helmholtz free energy density for electrons and ions,
the \(\vec{E} \times \vec{B}\) energy density, the parallel energy densities for electrons and ions and the perturbed magnetic field energy density.
In \(\Lambda\) we insert the dissipative terms of Section~\ref{sec:dissres} and use $\Lambda_U := \Lambda_{mNU}/mU - U \Lambda_N/ N$.

The dissipation term can be further simplified to
\begin{align}
    \Lambda_{\mathcal E} =&  -\sum_s \nc\left[\left( T\left( 1+\ln{(N/n_{e0})}\right) + q\psi + \frac{1}{2} mU_\parallel^2 \right)
  \vec j_{N,\nu} \right]
   \nonumber\\
  & - \nc ( U_\parallel \bar{ \vec j}_{mNU,\nu})
    \nonumber\\
  &+ \bar{ \vec j}_{mNU,\nu} \cn U_\parallel 
   + \vec j_{N,\nu} \cn( \ln N / n_{e0} - q\psi ) 
\end{align}
where we use $\nc\bar{ \vec j}_{mNU,\nu} :=  \mu_{U,\perp}(-\Delta_\perp)^2 u_{\parallel,e}  - \mu_{\parallel,e} \Delta_\parallel u_e $. 
The dissipation term thus consists of a diffusive energy
current under a total divergence and a dissipation contribution. Focusing on the parallel diffusion terms
we find for the dissipative contribution:
\begin{align}
     \bar{ \vec j}_{mNU,\nu} \cn U_\parallel 
   + \vec j_{N,\nu} \cn( \ln (N / n_{e0}) - q\psi ) = 
   \nonumber\\
   -\mu_{\parallel,U} (\npar U)^2 - \mu_{\parallel,N} \frac{(\npar N)^2}{N} - q \mu_{\parallel,N} \npar N \npar \psi
\end{align}
The first two terms are always negative and thus always dissipate energy. The last term containing the potential vanishes under species summation at least to zeroth order with $n_e\approx N_i$ and $\psi_i \approx \phi$. 

The term $R_{\mathcal E}$ is approximately quadratic in the sense that $R_{\mathcal E} \approx - \eta_\parallel j_\parallel^2$, which is the familiar Joule heating term. Since
we have an isothermal model this term appears as an energy dissipation term. 
The source term $S_{\mathcal E}$ dissipates parallel kinetic energy $-0.5mU_\parallel^2 S_N <0$ but generates free energy $\ln N S_N > 0$.

\begin{figure*}[htbp]
    \centering
    \includegraphics[width = \textwidth]{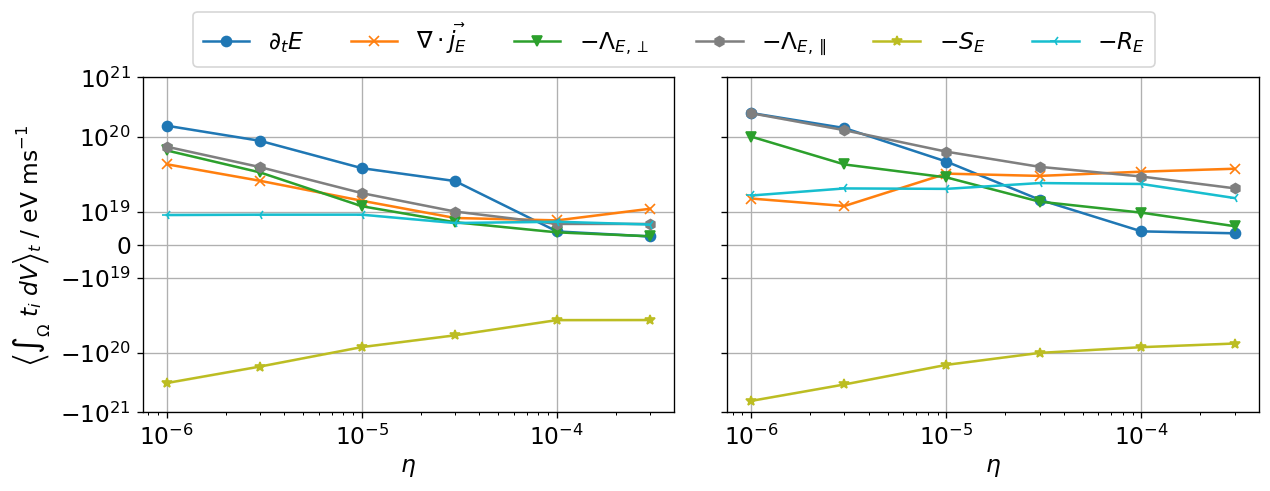}
    \caption{Energy conservation equation \eqref{eq:energy_conservation}: the terms Eq.~\eqref{eq:definition_ti} with error bar Eq.~\eqref{eq:error-ti} for $T_i=0$ (left) and $T_i=T_e$ (right).  The error bars are too small to be visible in the plot and are separately shown in Fig.~\ref{fig:relative-errors}.}
    \label{fig:energy-importance}
\end{figure*}

The integration region in time remains unchanged and we can compute the time and volume integrated terms
Eq.~\eqref{eq:definition_ti} with error bar Eq.~\eqref{eq:error-ti} in Fig.~\ref{fig:energy-importance}.
The relative errors of the terms must again be read from Fig.~\ref{fig:relative-errors} and are below $1$\% for all simulations. The global relative error in energy is generally a factor $2-5$ smaller than the error in mass.

In Fig.~\ref{fig:energy-importance} we see that the energy source $S_{\mathcal E}$ is the largest (and only) negative contributor in the equation. From Eq.~\eqref{eq:energy-source} we see that it is in fact the density source $S_{n_e}$ that translates to a source of energy. The magnitude of the energy source decreases by approximately a factor $10$ from smallest to highest resistivity. Since the density source does not vary much in Fig.~\ref{fig:mass-importance}, this is likely a simple consequence of the decreasing electron temperature in our parameter scan in Table~\ref{tab:parameters}. The energy source is balanced by the energy flux out of the last closed flux surface $\vec j_{\mathcal E}$, the parallel energy dissipation $\Lambda_{\mathcal E,\parallel}$, the Joule heat $R_{\mathcal E}$, the perpendicular energy dissipation $\Lambda_{\mathcal E,\perp}$ and the energy gain $\partial_t \mathcal E$. Few clear trends with resistivity can be inferred from the plot. The parallel energy dissipation is systematically larger than the perpendicular energy dissipation. The resistivity term $R_{\mathcal E}$ becomes relatively less important for smaller resistivities $\eta$ than for higher resistivities. For $T_i=0$ the energy gain $\partial_t \mathcal E$ is most important for small resistivities $\eta < 10^{-4}$ but least important else.  
The energy flux term $\vec j_{\mathcal E}$ is most important for $\eta \geq 10^{-4}$ but small compared to the other terms for $\eta < 10^{-5}$. 
\subsubsection{Parallel momentum balance} \label{sec:parallel-mom}
In the parallel momentum equation~\eqref{eq:parallel-mom} for ions we insert the mirror force Eq.~\eqref{eq:parallel-mirror} and use $-(\bhat+\bperp)\cn \ln B = \nc ( \bhat+\bperp) $ to get
\begin{align}
\frac{\partial}{\partial t} &\left(m_iN_i U_{\parallel,i}\right) + eN_i \frac{\partial}{\partial t} A_\parallel 
+ \nc \vec J_{mNU,i} \nonumber \\
& +T_i (\bhat + \bperp)\cn N_i + \frac{m_i}{e} N_i U_{\parallel,i} T_i \KK\cn\ln B \nonumber\\
&- F_{mNU,\psi}  + R_{\parallel,e} -  \Lambda_{mNU,i}=0,
   \label{eq:parallel-ion-momentum}
\end{align}
with ion momentum current
\begin{align}\label{eq:parallel-ion-mom-current}
\vec J_{mNU,i} :=&  \vec j_\parallel + \vec j_A + \vec j_E + \vec j_{C} ,\\
 \vec j_{mNU,\parallel} :=& m_iN_iU_{\parallel,i}^2 \bhat , \nonumber\\
 \vec j_{mNU,A} :=&  m_iN_iU_{\parallel,i}^2\bperp , \nonumber\\
\vec j_{mNU,E} :=& m_iN_i U_{\parallel,i}\frac{\bhat\times  \vn\psi }{B} ,
 \nonumber\\
    \vec j_{mNU,C} :=& \frac{m_i}{e}U_{\parallel,i} N_i \left(3T_i + m_iU_{\parallel,i}^2 \right)\vec \KK \nonumber\\
    &+\frac{m_i}{e}U_{\parallel,i} N_iT_i \KB, \nonumber
\end{align}
as well as resistivity term and the parallel electric force
\begin{align}
    R_{\parallel,e}:=& \eta_\parallel e^2 n_e ( N_i U_{\parallel,i} - n_e u_{\parallel,e}) , \\
    F_{mNU,\psi} =& -eN_i(\bhat +
    \bperp) \cdot \vn\psi \nonumber\\
    &-m_i N_i U_{\parallel,i}\KK\cdot\vn \psi .
    \label{eq:parallel-ion-electric}
\end{align} 
Note that the total divergences $\nc \vec j_{mNU,\parallel}$ and $\Lambda_{mNU,\parallel}$, parallel flux and viscosity terms, again vanish exactly under the flux-surface average. 
\begin{figure*}[htbp]
    \centering
    \includegraphics[width = \textwidth]{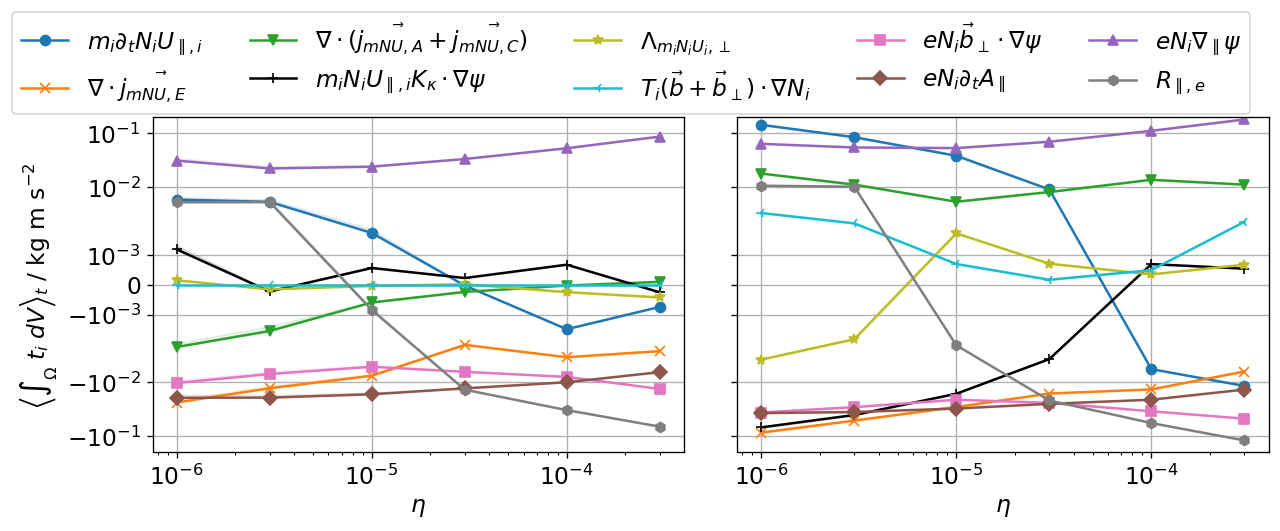}
    \includegraphics[width = \textwidth]{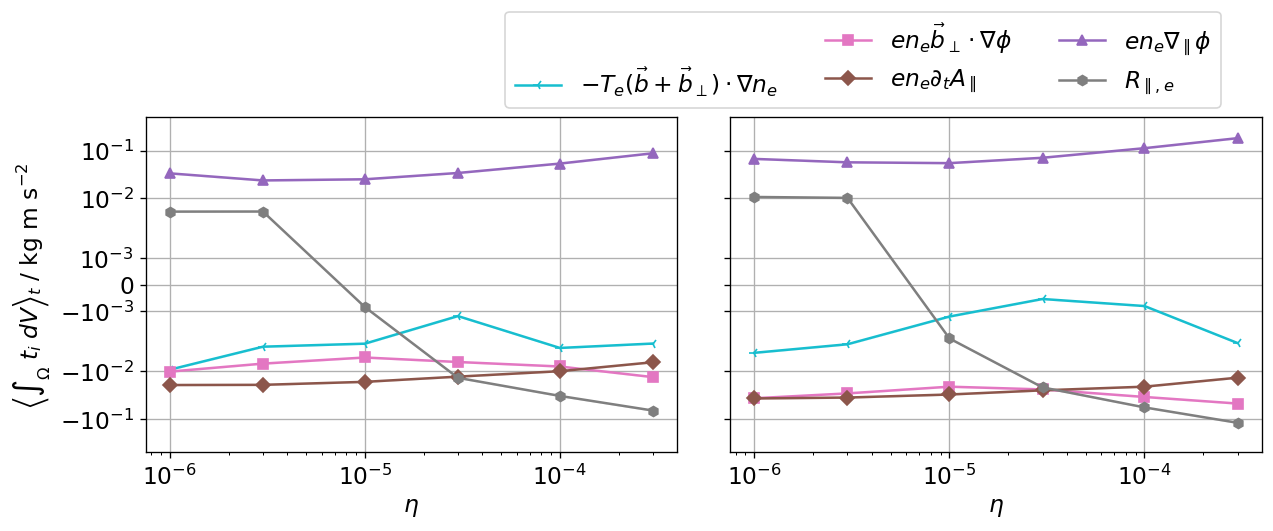}
    \caption{The parallel momentum balance (top) Eq.~\eqref{eq:parallel-ion-momentum}
    and the parallel electron force balance (bottom) Eq.~\eqref{eq:parallel-force-balance}:  the terms Eq.~\eqref{eq:definition_ti} with error bar Eq.~\eqref{eq:error-ti} for $T_i=0$ (left) and $T_i=T_e$ (right). The error bars are too small to be visible in the plot and are separately shown in Fig.~\ref{fig:relative-errors}.
    }
    \label{fig:niui-importance}
\end{figure*}
We plot the terms of the ion momentum equation in the top half of Fig.~\ref{fig:niui-importance}.
Again, the error bars are invisible and are separately plotted in Fig.~\ref{fig:relative-errors}.
There we find relative errors for $T_i=T_e$ between $10^{-3}$ and $3\cdot 10^{-2}$. Each term of the ion momentum equation thus has a relative error of maximal $3$\%. This is true also for $T_e=0$ and $\eta>10^{-5}$ simulations. However, for $T_i=0$ and $\eta\leq 10^{-5}$ the relative error climbs to about $10$\%. This can be reasoned in the smallness of the terms in Fig.~\ref{fig:niui-importance}, i.e. the absolute error of the equation remains the same across simulations but the term $\sum_i |T_i^\text{num}|$ in Eq.~\eqref{eq:rel-global-error} is small for $T_i=0$ and small $\eta$. 

In Fig.~\ref{fig:niui-importance} the largest positive term is the parallel electric force $eN_i\npar \psi$. To this add negative contributions from the gauge term $eN_i\partial_t A_\parallel$ and the magnetic flutter $eN_i\bperp\cn\psi$. The resistivity term $R_{\parallel,e}$, as expected, 
makes a significant contribution only for large $\eta>10^{-5}$ for both $T_i=0$ as well as $T_i=T_e$.
The \ExB{} flux is the final significant term and decreases in magnitude with $\eta$. The absolute value
is however larger for $T_i=T_e$ than for $T_i=0$. 

For $T_i=T_e$ and small resistivities the term $m_i\partial_t N_i U_{\parallel,i}$ is the largest positive term. This indicates positive acceleration, while for large resistivites $\eta > 3\cdot 10^{-5}$ there is acceleration in the opposite direction. For $T_i=0$ the same trend can be observed, however, the magnitude of the term is about a factor $10$ smaller than for the $T_i=T_e$ simulations. We will discuss this further in Section~\ref{sec:acceleration}.

\subsubsection{Parallel electron force balance} \label{sec:parallel-force}
The parallel electron momentum equation is given by Eq.~\eqref{eq:parallel-ion-momentum}
with electron instead of ion labels. In a plot of the terms analogous to the ion momentum plot Fig.~\ref{fig:niui-importance} (top) it turns out that most of the terms are very close to zero.
We thus gather only the dominant terms in the electron momentum equation neglecting all terms proportional to the electron mass with $m_e=0$. This leaves the parallel force balance
\begin{align}\label{eq:parallel-force-balance}
    &-T_e(\bhat + \bperp) \cn n_e 
     \nonumber\\
    &+en_e\left( \left( \bhat + \bperp \right) \cn \phi + \frac{\partial A_\parallel}{\partial t} \right)
    \nonumber\\
   & +R_{\parallel,e} \approx 0
\end{align}
In the bottom half of Fig.~\ref{fig:niui-importance} we plot the terms of the parallel force balance. The
relative global error of this equation is generally the smallest among all the equations that we test. In Fig.~\ref{fig:relative-errors} we see that the error is of excellent orders $10^{-4}$ and $10^{-3}$, which lies in the range of the value for $m_e/m_i = 2.7 \cdot10^{-4}$. This confirms that at least under volume integration Eq.~\eqref{eq:parallel-force-balance} is very well fulfilled even if it is not analytically exact.

Analogous to the ion momentum equation the largest term in the electron force balance is the parallel electric force $en_e \npar \phi$.
Notice here that the colours of Fig.~\ref{fig:niui-importance} (top) and \ref{fig:niui-importance} (bottom) coincide for analogous terms. 
In fact, visually the terms $en_e\npar \phi$, $en_e\bperp\cn \phi$ and $en_e \partial_t A_\parallel$, i.e. all terms of the electric field are indistinguishable from $eN_i \partial_t A_\parallel$, $eN_i\bperp\cn\psi$ and $eN_i\npar\psi$. We will use this to further study the total momentum equation in Section~\ref{sec:acceleration}. 

\subsection{Parallel Acceleration}\label{sec:acceleration}
\begin{figure*}[htbp]
    \centering
    \includegraphics[width = \textwidth]{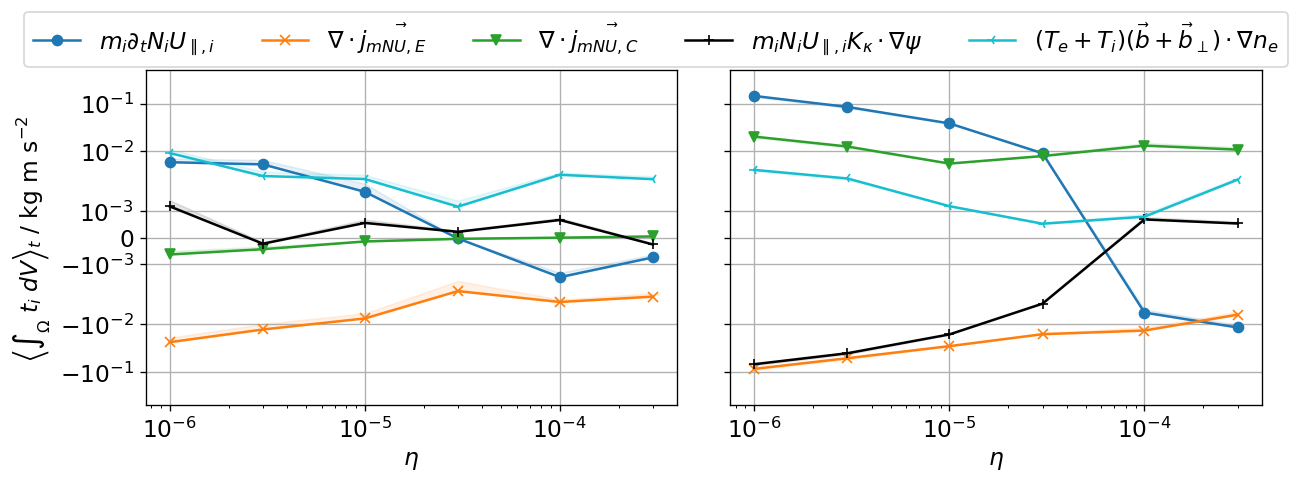}
    \caption{The sum of electron force balance and the parallel ion momentum equation (Fig.~\ref{fig:niui-importance}) neglecting small terms. The summed electric force is close to zero and drops out as does the resistivity. The error bars in the $T_i=0$ (left) plot become visible for $\eta\leq 3\cdot 10^{-5}$ while staying invisible for $T_i=T_e$ (right).
    }
    \label{fig:sum-ue-ui-importance}
\end{figure*}

Fig.~\ref{fig:niui-importance} is visually overburdened due to the number of displayed terms and thus hard to physically interpret further. Thus, we here simplify the discussion by focusing on the total momentum balance. First, we see in Fig.~\ref{fig:niui-importance} that the electron and ion components of the electric field and the resistivity are visually equal. Neglecting those terms we sum the ion and electron momentum equations to get
\begin{align}
    m_i\frac{\partial}{\partial t} &N_iU_{\parallel,i} + \nc (\vec j_{mNU,E}+\vec j_{mNU,C}) \nonumber\\
   & + m_i N_iU_{\parallel,i}\KK \cn \psi + (T_e+T_i)(\bhat+\bperp)\cn n_e
   \label{eq:sum-ue-ui}
\end{align}
We further neglect the term $\nc \vec j_{mNU,A}$ and $\Lambda_{mNU,\perp}$ and approximate $T_i(\bhat + \bperp)\cn N_i \approx T_i(\bhat + \bperp)\cn n_e$. The result is shown in Fig.~\ref{fig:sum-ue-ui-importance}.

The error bars in Fig.~\ref{fig:sum-ue-ui-importance} are visible in particular in the $T_i=0$ plot, however the plot is easier to interpret than Fig.~\ref{fig:niui-importance}. We now clearly see the positive acceleration in the $T_i=T_e$ plot for $\eta \leq 10^{-4}$. For $\eta\geq 10^{-4}$ the parallel acceleration is negative. The $T_i=0$ plot shows the same trends but the acceleration is more than a factor $10$ smaller than for $T_i=T_e$. 

Four candidates explain the observed accelerations. The \ExB{} flux of parallel momentum is negative signifying that positive momentum is lost to the plasma (or negative momentum enters the plasma) via the radial transport. The \ExB{} flux decreases in magnitude with $\eta$ for both $T_i=0$ and $T_i=T_e$ but is about a factor $2-4$ larger for $T_i=T_e$ than for $T_i=0$.  For $T_i=0$ the two terms $\nc \vec j_{mNU,C}$ and $m_iN_iU_{\parallel,i}\KK\cn \psi$ are close to zero for all $\eta$. The only remaining term for $T_i=0$ is thus
the parallel gradient $T_e(\bhat+\bperp) \cn n_e$, which remains roughly constant in $\eta$. 

For $T_e=T_i$ the term $(T_e+T_i)(\bhat+\bperp) \cn n_e$ is positive but much smaller than the curvature contribution. The second curvature term $\nc m_iN_iU_{\parallel,i}\KK\cn \psi$ is strongly negative for $\eta < 10^{-4}$ but jumps to a positive contribution at $\eta = 10^{-4}$ thus facilitating the associated negative acceleration. 
The term $\nc \vec j_{mNU,C}$ in Fig.~\ref{fig:sum-ue-ui-importance} represents the total flux of ion momentum through the last closed flux surface by curvature drifts, while the term $m_iN_iU_{\parallel,i}\KK\cn \psi$ appears as a drift correction to the parallel electric force term~\eqref{eq:parallel-electric}. In our previous theoretical analysis both curvature terms were neglected as small~\cite{wiesenberger2020angular} but for $T_i=T_e$ each term has similar contribution in magnitude to the radial \ExB{} momentum flux.


\subsection{Mass and energy confinement times} \label{sec:confinement}
\begin{figure}[htbp]
    \centering
    \includegraphics[width = 0.48\textwidth]{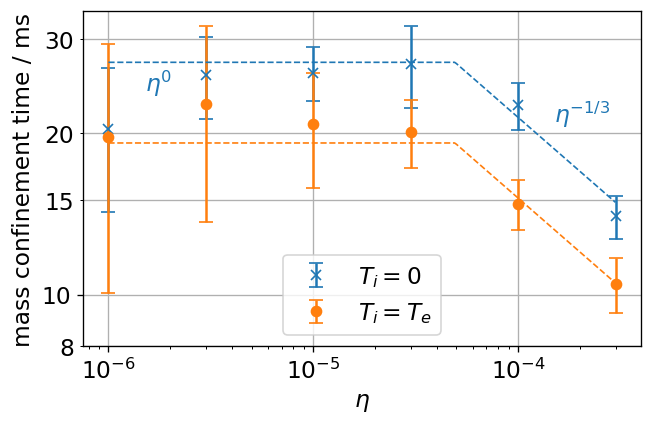}
    \caption{ The mass confinement times $\tau_M$ Eq.~\eqref{eq:mass-confinement-time}. The fit is given by Eq.~\eqref{eq:mass-scaling}.
    }
    \label{fig:mass-confinement-times}
\end{figure}

\begin{figure}[htbp]
    \centering
    \includegraphics[width = 0.48\textwidth]{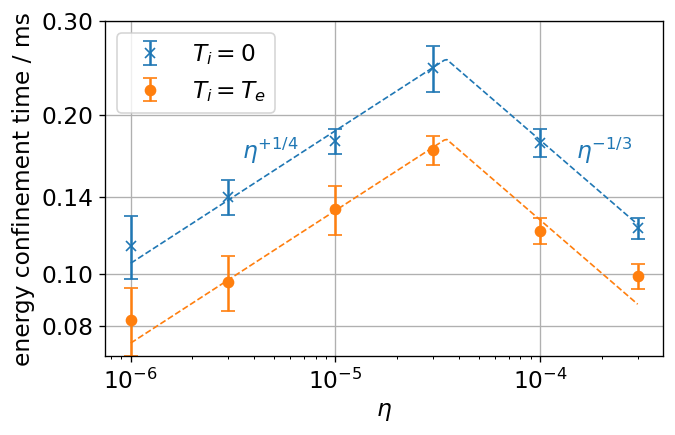}
    \caption{The energy confinement times $\tau_E $ Eq.~\eqref{eq:energy-confinement-time}. The fit is given by Eq.~\eqref{eq:energy-scaling}.
    }
    \label{fig:energy-confinement-times}
\end{figure}

From our analysis of the mass conservation equation in Fig.~\ref{fig:mass-importance} and the energy conservation equation in Fig.~\ref{fig:energy-importance} it is straightforward to extract confinement times. As we explained before the volume integral of $\nc \vec j$ yields the total flux out of the closed fieldline region $\int \vec j \cdot \vec \dA  $. We thus start with the definition 
of the total particle number and energy within the confined region
\begin{align}
    M(t) = \int_\Omega n_e \dV \\
    E(t) = \int_\Omega \mathcal E \dV
\end{align}
We can then compare these with the total loss of particles and energy.
The particle loss is simply the total flux $\vec j_{n_e}$ (Eq.~\eqref{eq:mass-flux}) integrated over the last closed flux surface.
We can neglect the diffusive transport from Fig.~\ref{fig:mass-importance} as close to zero.
The losses of energy consist of the energy flux
out of the last closed flux surface, but also of the energy dissipation through
diffusion and the resistivity. 
We thus define
\begin{align}
    \tau_M :=& \frac{\langle M\rangle_t }{\left\langle \int_\text{LCFS} \vec j_{n_e} \cdot \vec\dA\right\rangle_t} \label{eq:mass-confinement-time}\\
    \tau_E :=& \frac{\langle E\rangle_t }{\left\langle \int_\text{LCFS} \vec j_{\mathcal E}\cdot \vec \dA - \int_\Omega (\Lambda_\mathcal E + R_{\mathcal E})\dV\right\rangle_t}  \label{eq:energy-confinement-time}
\end{align}
In Fig.~\ref{fig:mass-confinement-times} and \ref{fig:energy-confinement-times}
we present the resulting values for our simulations. Note that the total particle number $\RA{M}_t = (2.3\pm0.1)\cdot 10^{19}$ is roughly constant for all simulations.
The error bars are computed from the fluctuation amplitudes of all quantities in Eqs.~\eqref{eq:mass-confinement-time} and \eqref{eq:energy-confinement-time}. The relative numerical errors are negligible at $1\%$ as established in Section~\ref{sec:verification}.
Two regimes are visible in both plots with a transition at $\eta_\text{crit}\approx5\cdot 10^{-5}$
for both $T_i=0$ as well as $T_i=T_e$.

The mass confinement times in Fig.~\ref{fig:mass-confinement-times} reach roughly constant values for $\eta<3\cdot 10^{-5}$ while for $\eta>10^{-5}$ there is a decrease
of confinement with increasing resistivity. The drop in mass confinement above the critical $\eta$ could be related to the discussion of the density limit~\cite{Eich2020,Giacomin2022regimes} in the operational space of tokamaks. The constant regime should be regarded tentatively as the fluctuations are particularly large in this regime, especially for $T_i=T_e$.
The values for $T_i=0$ are a factor $\sqrt{1+T_i/T_e}$ larger than the ones for $T_i=T_e$ within the error bars. 
We can tentatively fit a power law of 
\begin{align}\label{eq:mass-scaling}
    \tau_M = \frac{c_M(n_0,\rho_s)}{\sqrt{1+T_i/T_e}}\begin{cases}
        1 &\text{ for } \eta<5\cdot 10^{-5} \\
        \eta^{-1/3} &\text{ for } \eta>5\cdot 10^{-5}
    \end{cases}
\end{align}
where $c_M(n_0,\rho_s)$ signifies the unknown dependency on the parameters $n_0$ and $\rho_s$ that we kept constant during our parameter scan. We remind the reader here that the values for both $T_e$ and $B_0$ decrease for increasing $\eta$ in our parameter scan as seen in Table~\ref{tab:parameters}.

For the energy we see a clear maximum in the confinement time at $\eta=3\cdot 10^{-5}$. The fluctuations are systematically smaller for the energy
confinement times than for the particle confinement times. However, the energy
confinement times are also approximately a factor $100$ smaller than the mass confinement times. This may be due to the fact that we have an isothermal model where Joule heat is not converted to an increase in temperature and is instead lost to the system. A tentative fit reveals 
\begin{align}\label{eq:energy-scaling}
    \tau_E = \frac{c_E(n_0,\rho_s)}{\sqrt{1+T_i/T_e}}\begin{cases}
    \eta^{+1/4}\text{ for }\eta<3.5\cdot 10^{-5} \\
        \eta^{-1/3} \text{ for } \eta>3.5\cdot 10^{-5}
    \end{cases}
\end{align}
where similar to Eq.~\eqref{eq:mass-scaling} the factor $c_E(n_0,\rho_s)$
encapsulates a yet unknown dependence on the parameters $n_0$ and $\rho_s$. 

The existence of a critical value for the plasma resistivity $\eta_{crit}\approx 5\cdot 10^{-5}$ for both mass and energy confinement points towards two different turbulent regimes above and below the critical value. 
Various candidates are discussed in the literature with the most likely ones being drift-wave turbulence for small $\eta$ and resistive ballooning type turbulence for high $\eta$~\cite{Zeiler1996,Scott2005,Giacomin2022regimes}. 
According to Reference~\cite{Scott2005} the transition between the two regimes happens at the resistive ballooning threshold at $\alpha_{t,crit} = 1$ with turbulence parameter $\alpha_t := \eta q^2R_0/\rho_s \approx 5\cdot 10^3 \eta$. With $\eta_{crit} = 5\cdot 10^{-5}$ we obtain $\alpha_{t,\text{crit, num}} \approx 0.25$, which is only a factor $4$ away from the theoretical prediction. 
The difference may be explained by geometrical factors like the presence of the X-point.

There is an apparent discrepancy in this explanation however, insofar the transport in drift-wave turbulence reduces for small $\eta$ (converging to the adiabatic case) and thus the confinement time should increase for decreasing $\eta$ instead of remaining constant. An explanation for the observed plateau in the mass confinement time could be so-called reactive instabilities, which are independent of $\eta$ and are due to a finite electron inertia~\cite{ScottBook}. Reactive instabilities are unphysical insofar they are an artefact of an isothermal gyro-fluid model and have no gyro-kinetic counterpart where Landau damping counteracts the effect of electron inertia. Note that this does not contradict Fig.~\ref{fig:niui-importance} where the electron inertia effect vanishes under volume integration. Locally, the electron inertia may still be important.

\section{Conclusion} \label{sec:conclusion}
We present a new version of the three-dimensional gyro-fluid turbulence code \feltor{}. $12$ simulations covering several milliseconds with different values for plasma resistivity and ion temperature and fixed values for plasma density and gyro-radius are setup, analysed and discussed. An efficient implementation on GPUs allows for simulation runtimes of about $1$ week per simulation. \feltor{} is verified using volume and time integrated conservation laws, mass, energy, momentum and force balance. Relative errors are generally below $1$\% for energy conservation and force balance while for mass and momentum conservation the errors climb to about $3$\% as seen in Fig.~\ref{fig:relative-errors}.  Only in the ion momentum balance and for vanishing ion temperature and small resistivity do we see relative errors of about $10$\%, which is reasoned in the smallness of the parallel acceleration compared to $T_i=T_e$ simulations with at the same time equal absolute errors. 

We systematically investigate the importance of the terms in the parallel momentum generation where we find that for increasing resistivity the direction of acceleration is swapped. This is caused mainly by an interplay of decreasing \ExB{} momentum transport and curvature drifts across the separatrix. 
The analysis of the momentum density $m_iN_iU_{\parallel,i}$ is related to intrinsic toroidal rotation in tokamaks and the angular momentum density $m_iN_iU_{\parallel,i}R$~\cite{Stoltzfus2019,wiesenberger2020angular}. A detailed analysis of rotation profiles and the angular momentum balance is here postponed to future analysis.  

Similar transitions from a low resistivity regime to a high resistivity regime happen for the mass and energy confinement times.
 Beyond the critical resistivity the mass and energy confinement decrease with increasing resistivity. Below it, the mass confinement remains roughly constant, while the energy confinement decreases with decreasing resistivity. This behaviour could be explained by so-called reactive instabilities, which are an artefact of electron inertia in isothermal gyro-fluid models and have no gyro-kinetic counterpart. A dynamic electron temperature should help counteract this effect in future works.
 The transition from drift-wave turbulence to resistive ballooning roughly coincides with the value predicted by the literature. Further parameter studies in $\rho_s$ and $n_0$ need to clarify the unknown dependence factors $c_M(n_0,\rho_s)$ and $c_E(n_0,\rho_s)$ in the observed scaling laws for $\tau_M (\eta)$~\eqref{eq:mass-scaling} and $\tau_E(\eta)$~\eqref{eq:energy-scaling}.

The capability of running numerically stable simulations for a set of different parameters with \feltor{} is an important milestone. We offer a first high level analysis of the run simulations and quantify numerical errors, leaving many questions open for future work as outlined above.
Furthermore, various physical model improvements can be added fairly straightforwardly within the \feltor{} framework. These include for example, dynamic temperature equations~\cite{Held2016blobs}, plasma-neutral collisions~\cite{Wiesenberger2022LWL}, arbitrary order polarisation terms~\cite{Held2020Pade,Held2023NOBblobs} and more.
\section*{Acknowledgements}
We thank A. Kendl for fruitful discussions.
This work has been carried out within the framework of the EUROfusion Consortium, funded by the European Union via the Euratom Research and Training Programme (Grant Agreement No 101052200 — EUROfusion). Views and opinions expressed are however those of the author(s) only and do not necessarily reflect those of the European Union or the European Commission. Neither the European Union nor the European Commission can be held responsible for them. 
This work was supported by the UiT Aurora Centre Program, UiT The Arctic University of Norway (2020).
This research was funded in whole or in part by the Austrian Science Fund (FWF) [P 34241-N]. 
For the purpose of Open Access, the author has applied a CC BY public copyright license to any Author Accepted Manuscript (AAM) version arising from this submission.
This work was supported by a research grant (15483) from VILLUM Fonden, Denmark.
\appendix

\section{General magnetic field expressions} \label{sec:formulary}
We assume a three-dimensional flat space with arbitrary coordinate
system $\vec x :=\{x_0, x_1, x_2\}$, metric
tensor $g$ and volume element $\sqrt{g} := \sqrt{\det g}$.
Given a vector field $\vec B(\vec x)$ with unit vector $\bhat(\vec x) := (\vec B/B)({\vec x})$
we can define various differential operations in table~\ref{tab:operators}.
\begin{table}[htbp]
\caption{Definitions of geometric quantities for given vector field $\vec B(\vec x)$, unit vector field $\bhat := \vec B/B$ and metric tensor $g$.
}\label{tab:operators}
\begin{tabular}{ll}
\rowcolor{gray!50}\textbf{Name} &  \textbf{Definition} \\
\midrule
    Projection Tensor&
    $h := g - \bhat\bhat$ \\
    Perp. gradient&
    $ \nperp f := \bhat\times(\vn f\times \bhat ) \equiv
    h \cdot \vn f$ \\
    Perp. divergence&
    $ \nperp^\dagger \cdot \vec v := -\nc( h \cdot \vec v) = -\nc\vec v_\perp$ \\
    Perp. Laplacian &
    $ \Delta_\perp f:= \nc (\nperp f)
    = \nc( h\cdot\vn f) $  \\
    Curl-b Curvature &
    $\KK := \frac{1}{B}(\vn \times \bhat)$ \\[4pt]
    Grad-B Curvature &
    $ \KB := \frac{1}{B}(\bhat \times\vn \ln B)$ \\[4pt]
    Curvature &
    $\vec{ K} :=
     \vn \times \frac{\bhat}{B} $,\\[4pt]
    Parallel derivative&
    $ \npar f := \bhat\cdot\vn f$ \\
     Parallel Laplacian&
     $\Delta_\parallel f:= \vec{\vn} \cdot ( \bhat\bhat\cdot\vec{\vn} f )$\\
\bottomrule
\end{tabular}
\end{table}
Explicit expressions for these expressions
depend on the choice of the magnetic field and the underlying coordinate system.
Note that we have
\begingroup
\allowdisplaybreaks
\begin{align}
h^2 &= h, \\
    \nc \KK
= -\nc \KB &= -\KK\cn\ln B, \\
    \nc\vec{ K} &= 0, \\
    \vec K &= \KB + \KK \\
    \KK - \KB &= \frac{1}{B^2} (\vn \times \vec B), \\
    \vec{\vn}\cdot\left(\frac{\bhat\times\vec{\vn} f}{B}\right) &=\vec K \cn f, \\
    \Delta_\perp f&= -\nperp^\dagger\cdot\nperp f , \\
    \npar \ln B &= -\vec\nc\bhat .
    \label{eq:curl_curvature}
\end{align}
\endgroup
The last equality holds with $\vec\nc \vec B = 0$.
Furthermore, we have
\begin{align}
\bhat \cdot \left(\vn f \times\vn g\right) &=
    b_i \varepsilon^{ijk}\partial_j f\partial_k g/\sqrt{g},
\end{align}
In any arbitrary coordinate system we have
\begin{align}
(\vn f)^i = g^{ij}\partial_j f , \qquad
\nc \vec v = \frac{1}{\sqrt{g}}\partial_i \left(\sqrt{g} v^i\right) ,\nonumber\\
(\vec v \times \vec w)^i = \frac{1}{\sqrt{g}}\varepsilon^{ijk} v_jw_k .
\end{align}
 with $b^i$ the contra- and $b_i$ the co-variant components of $\bhat$,
$\eps^{ijk}$ the Levi-Civita symbols and $g^{ij}$ the contra-variant elements of the metric tensor.
\section{Data access} \label{app:data}
The \feltor{} code is available freely on GitHub at \url{https://github.com/feltor-dev/feltor} with the latest release tracked on Zenodo~\cite{Feltor}. It includes the dg library and the three-dimensional code used for this paper. The magnetic field equilibrium, wall and sheath domains and simulation box are setup using our
earlier mentioned
\url{https://github.com/feltor-dev/magneticfielddb} Python repository. The parameter
scan is setup using \url{https://github.com/mwiesenberger/feltorutilities} which in turn is based on the \texttt{simplesimdb} Python package developed at \url{https://github.com/mwiesenberger/simplesimdb}. \texttt{Simplesimdb} is a free simulation database manager in Python that allows to run / submit, access and manage simulations using a unified Python interface. In order to help analyse the simulation data in Python we use \texttt{xFeltor} \url{https://github.com/feltor-dev/xFELTOR}, an interface to the \texttt{xarray} Python package and \texttt{pyFeltor}  \url{https://github.com/feltor-dev/pyFeltor}, an implementation of basic dG numerical methods in Python.  All three-dimensional renderings were setup in ParaView~\cite{paraview} the remaining analysis is available as Jupyter Notebooks in \url{https://github.com/mwiesenberger/data-analysis-3d}.

\bibliography{references}
\bibliographystyle{iopart-num.bst}


\end{document}